\documentclass{aa}
\usepackage{graphicx}
\begin{document}

  \title{ The Luminosity-Metallicity Relation of distant luminous
infrared galaxies 
   \thanks {Based on observations collected with the ESO Very Large
   Telescope at the Paranal Observatory (66.A-0599(A)) 
   and with ISOCAM Deep Survey and ISO-CFRS follow-up}} 

  \titlerunning{ The L-Z relation of distant luminous  infrared galaxies  }

   \author{Y. C. Liang,
           \inst{1,2}
	F. Hammer
	   \inst{1},	 	  
	H. Flores 
	  \inst{1},   
	D. Elbaz
          \inst{3},
	D. Marcillac
	  \inst{3},
	C. J. Cesarsky
	   \inst{4}     
          }
   \authorrunning{Liang et al.} 

   \offprints{Y. C. Liang,
              email: ycliang @bao.ac.cn; F. Hammer, email: francois.hammer@obspm.fr
          }
  \institute{GEPI, Observatoire de Paris, Section de Meudon, 92195 Meudon Cedex, France 
 \and
             National Astronomical Observatories, Chinese Academy of Sciences, No.20A Datun Road,
	     Chaoyang District, Beijing 100012, P. R. China         	      
 \and
             CEA, Saclay-Service d'Astrophysique, Orme des Merisiers, F91191
	     Gif-sur-Yvette Cedex, France 
\and
              ESO, Karl-Schwarzschild Strase 2, D85748 Garching bei Munchen, Germany 
              }	      
  
   \date{Received 25 Novermber 2003; accepted 26 April 2004}

\abstract{ 
One hundred and five 15$\mu m$ selected objects 
in three ISO ($Infrared~ Space~ Observatory$) deep survey fields (CFRS 3$^{h}$, UDSR and UDSF)
are studied on the basis of their high quality optical spectra with resolution R$>$1000 
from VLT/FORS2.  
Ninety two objects (88\%) have secure redshifts, ranging from 0 to 1.16 with 
a median value of $z_{\rm med}$=0.587. 
\newline
Considerable care is taken in estimating the extinction property of individual
galaxy, which can seriously affect diagnostic diagrams and estimates of 
star formation rates (SFRs) and of metal abundances. Two independent estimates of the extinction
have been made, e.g. Balmer line ratio and 
energy balance between infrared (IR) and H$\beta$ luminosities. 
For most of the sources, we find a good agreement between the two extinction
coefficients (within $\pm$0.64 rms in $A_{V}$, the extinction in V band), with median values of 
$A_V$(IR) = 2.36 and $A_V$(Balmer)= 1.82 for z$>$0.4 luminous IR galaxies (LIRGs). 
At z$>$0.4, our sample show 
many properties (IR luminosity, continuum color, ionization and extinction) 
strikingly in common with those of local
 (IRAS) LIRGs studied by Veilleux et al. (1995). 
Thus, our sample can provide a good
representation of LIRGs in the distant Universe.\newline
We confirm that most ($>$77\%) ISO 15$\mu m$ selected galaxies are dominated by star formation.
 Oxygen abundances in interstellar medium in the galaxies 
are estimated from the extinction corrected
``strong" emission line ratios (e.g. \ion{[O}{ii]}/H$\beta$, 
\ion{[O}{iii]}/H$\beta$ and \ion{[O}{iii]}/\ion{[O}{ii]}).
The derived 12+log(O/H) values range from 8.36 to 8.93 for the z$>$0.4 galaxies 
with a median value of 8.67.
Distant LIRGs present a metal content less than half of that of the local 
bright disks (i.e. $L^*$). 
Their properties
can be reproduced with infall models though one has to limit the infall time to avoid
overproduction of metals at late time. The models predict that total masses (gas + stars) of the distant LIRGs are
from $10^{11} M_{\odot}$ to $\le$ $10^{12} M_{\odot}$. 
A significant fraction of distant large disks are indeed LIRGs. 
Such massive disks could have formed $\sim$50\% of their metals and
 stellar masses since z$\sim$1.
} 

 \maketitle

%
%
\section{Introduction}

The IRAS all-sky survey
detected tens of thousands of galaxies
with far-infrared (far-IR) radiation luminosities from less than $10^{6} L_{\odot}$
to $\sim$$10^{13} L_{\odot}$ up to a moderate redshift ($z$$\sim$0.3).   
However, luminous infrared galaxies (LIRGs) are not typical of local galaxy population,
and they account for only $\sim$2\% of the local bolometric
luminosity density (Soifer et al. 1987; Sanders \& Mirabel 1996).

However, the COsmic Background Explorer (COBE) observations imply  that there
likely exists a very significant contribution of dust-obscured star formation
at high redshifts  (Puget et al. 1996; Genzel \& Cesarsky 2000). The ISO made
it possible to study the infrared emission of galaxies at $z \geq 0.5$, which 
plays an important role in understanding the co-moving star formation density 
evolution with look-back time. The ISO  mid infrared camera (ISOCAM) (Cesarsky
et al. 1996)  is $\sim$$10^{3}$ times more sensitive and has 60 times higher
spatial resolution than IRAS. The mid-IR ISOCAM 15 $\mu$m source counts provide
evidence for strong IR light density evolution, as revealed by the strong
excess of 15 $\mu$m counts above the predictions of  non-evolution models at
sub-mJy (Elbaz et al. 1999; Aussel et al. 1999). The cosmic infrared background
resolved by ISOCAM shows that the co-moving density of  infrared light due to
the luminous IR galaxies ($L_{\rm IR}\geq$10$^{11}L_{\odot}$)  was more than 40
times larger at $z\sim$1 than today (Elbaz et al. 2002).  The main driver for
this evolution is the  luminous infrared starburst galaxies seen by ISO at
$z$$>$0.4,  which form stars at a rate of more than $50M_{\odot}$\,yr$^{-1}$
(Flores et al. 1999). 

Based on the correlation analysis of deep X-ray and mid-IR observations in 
Lockman Hole and Hubble Deep Field North (HDF-N), Fadda et al. (2002)  found
that the active galactic nuclei (AGN) contribution to the 15 $\mu$m background
is only 17$\pm$2\%. They concluded that the population of IR luminous galaxies
detected in the ISOCAM deep surveys, and the cosmic infrared background sources
themselves,  are mostly dust-obscured starbursts (also see Elbaz et al. 2002).
Reviews of extragalactic results from ISO can be found in Genzel \&
Cesarsky (2000),  Franceschini et al. (2001) and Elbaz \& Cesarsky (2003).

Recently, Flores et al. (2004a) studied the interstellar extinction and SFRs of
16 luminous infrared galaxies in Canada-France Redshift Survey (CFRS) 3$^{h}$
and 14$^{h}$ fields using the spectra from the European Southern Observatory
(ESO) Very Large Telescope (VLT) and Canada-France-Hawaii Telescope (CFHT). 
They found that the extinction coefficients obtained from H$\gamma$/H$\beta$
(using the VLT/FORS2 or CFHT spectra)  and H$\alpha$/H$\beta$ (combining the
VLT/FORS2 and VLT/ISAAC spectra)   are in agreement, and that SFRs
derived from H$\alpha$ are consistent with those from infrared luminosities,
except for the galaxies near the  ultra luminous IR galaxy (ULIRG) regime
($L_{\rm IR}$ $>$ $10^{12}$ $L_{\odot}$).

Spectrophotometric properties of IRAS galaxies have been
studied in detail, providing a full diagnostic of their ISM properties 
(Veilleux et al. 1995, hereafter V95; Kim et al. 1995). 
However, at higher redshifts, studies of ISOCAM sources have
mostly focused on source counts and SFRs. Very few is known about chemical properties of distant LIRGs,
including about their metal content, and to fill this gap is the main objective of this paper.

In the local Universe, metallicity is well correlated with the absolute
luminosity (stellar mass) of galaxies over a wide magnitude range (e.g. 7-9
mag)  (Zaritsky et al. 1994; Richer \& McCall 1995; Telles \& Terlevich 1997;
Contini et al. 2002; Melbourne \& Salzer 2002; Lamareille et al. 2004). Some
results have been obtained on  the luminosity-metallicity (L-Z) relations in
the intermediate-redshift Universe. Kobulnicky \& Zaritsky (1999) found that
the L$-$Z  relations of 14 intermediate-$z$  emission line galaxies  with
$0.1<z<0.5$ are consistent with those   of the local spiral and irregular
galaxies   studied by Zaritsky et al. (1994), Telles \& Terlevich (1997) and 
Richer \& McCall (1995). The 16 CFRS galaxies at $z\sim$0.2 studied by Liang
et al. (2004)  fall well in the region occupied by the local spiral galaxies
(from Zaritsky et al. 1994).

At higher redshifts, Kobulnicky et al. (2003) have obtained the L$-$Z
relations of 64 galaxies from the Deep Groth Strip Survey (DGSS) which have
been separated into  three redshift ranges ($z$=0.2$-$0.4, 0.4$-$0.6, and
0.6$-$0.82).  In the highest redshift bin galaxies are brighter by $\sim$1 mag
relatively to those in the lowest redshift bin and  brighter by $\sim$2.4 mag
compared to the local ($z$$<$0.1) field galaxies  (from Kennicutt 1992a,b,
hereafter K92, and Jansen et al. 2000a,b, hereafter J20). Such a result is
confirmed by Maier et al. (2004).  These studies contrast with the results of
Lilly et al. (2003), who have found that the L$-$Z relation of 
most of their 66 CFRS
galaxies with $0.5<z<1$ is similar to that of the local galaxies from J20.
However, Lilly et al. (2003) have assumed a constant $A_{V}$=1 for accounting
for dust extinction. In this study, we investigate the L-Z relation for LIRGs
in $z>0.4$ Universe, after a detailed account for their dust extinction
properties.

This paper is organized as it follows.  In Sect.2,  we describe the sample
selection, the observations and the data reduction and analysis, while the 
redshift distribution and the spectrophotometric properties are presented in
Sect.3.  Sect. 2 and 3 are aiming at assessing whether our resulting sample
can be used to test the properties of distant LIRGs. Flux measurements,
interstellar extinction and SFRs  of the galaxies are shown in Sect.4. It
includes a detailed comparison between extinction parameter deduced from
Balmer line ratio to that derived from mid-IR luminosity.   In Sect.5, we
discuss the diagnostic diagrams to test the AGN contribution as well as the
ionization properties. In Sect.6, we present the luminosity-metallicity
relation (based on oxygen abundances) of distant LIRGs which is compared to
other galaxy samples.  Discussion and conclusion are given in Sections 7 and
8. Throughout this paper, a cosmological model with  $H_0$=70 km s$^{-1}$
Mpc$^{-1}$, $\Omega _M$=0.3 and $\Omega _\Lambda =0.7$ has been adopted.

\section{Sample selection, observations and data reduction}
\subsection{The fields}

Our sample galaxies were selected from three ISO deep survey fields: CFRS
3$^h$, Ultra-Deep-Survey-Rosat (UDSR) and  Ultra-Deep-Survey-FIRBACK (UDSF)
fields.

The CFRS was carried out in five moderate to high galactic latitude
($\mid$b$\mid$ $>$45$^{\circ}$) survey fields  of area
10$\arcmin$$\times$10$\arcmin$ chosen to match the field of view of the MOS
multiobject spectrograph on the 3.6 m CFHT. The 14$^{h}$ and 3$^{h}$ fields 
have been deeply imaged with the ISOCAM.  Combining the deep IR observation
and the deep optical and radio data, Flores et al. (1999) studied the 78
ISOCAM sources detected in the 14$^{h}$ field down to a 15 $\mu$m flux
$\geq$250$\mu$Jy. In the CFRS 3$^h$ field, 70 sources were detected, with the
15 $\mu$m fluxes in the range of 170$-$2100 $\mu$Jy (Flores et al. 2004a; 2004b,
in preparation). 

The UDSR field refers to the Marano field (centered at
$\alpha$(2000)=03$^h$15$^m$09$^s$, 
$\delta$(2000)=$-$55$^{\circ}$13$\arcmin$57$\arcsec$), which is a deep ROSAT
observation field. The deep ($\sim$80 or 120 ks integration time)  XMM-Newton
observations  (Giedke et al. 2001, Giedke et al. 2003) and the optical
identifications were also done (Lamer et al. 2003). FIRBACK is a deep survey
conducted with the ISOPHOT instrument aboard the ISO at an effective
wavelength of 175 $\mu$m.  The total survey covers more than 4 square degrees
located in one Southern and two Northern fields (Puget et al. 1999; Lagache \&
Dole 2001).  For the UDSR and UDSF fields, very deep ISOCAM follow-up  have
been done (Elbaz et al. 2004) reaching flux limits three times lower than for
the CFRS fields.

\subsection{The sample}

In the CFRS 3$^h$ field, we selected 25 targets out of the 70 ISO 15 $\mu$m
sources according to their ($\alpha$,$\delta$) positions to observe by using
VLT/FORS2.  Another 9 objects without 15 $\mu$m fluxes were added to these
observations to fill the FORS2 mask. In the UDSR field,   29 targets with ISO
15 $\mu$m fluxes were  selected to be observed by using VLT/FORS2. Another 6
non-ISO objects were also added. Three slits (number 18, 19 and 25) were
superposed to more than one object. In the UDSF field, 27 targets out of the
ISO sources list were selected to be observed by using VLT/FORS2. Another 9
objects without 15 $\mu$m fluxes were also selected.  Three slits (number 26,
27 and 29) were superposed to more than one object.

In total, 105 objects  were selected  for VLT/FORS2 spectral observations from
the three fields. The basic data of the target galaxies are given in Table 1
and 2.  The columns are the slit numbers (also CFRS name in Table 1), the 2000
epoch  coordinates, redshift $z$, I or R band photometric and spectral
magnitudes in the AB system,  aperture correction factor by comparing the  
photometric and spectral I or R band magnitudes, absolute B band magnitude
$M_B$ in the AB system, the spectral types of the objects,  and the related
infrared data including 15$\mu$m fluxes, far-IR luminosities and IR-SFRs.

\begin{figure}
\centering
\input epsf
\epsfverbosetrue
\epsfxsize 7.8cm 
\epsfbox{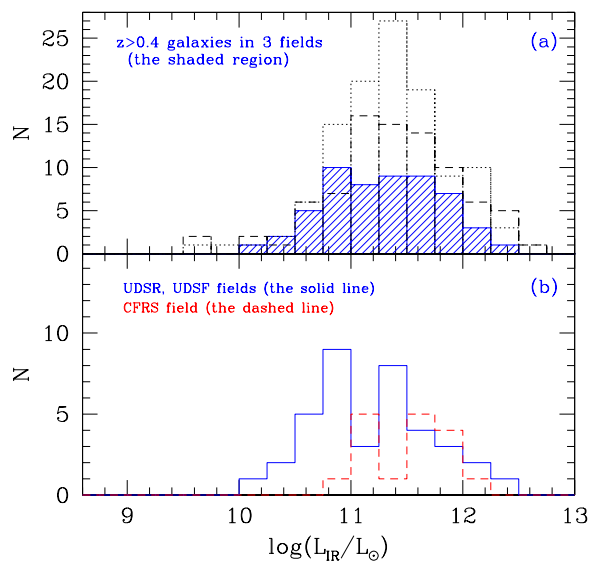}
\caption {{\bf (a)} The IR luminosity log($L_{\rm IR}/L_{\odot}$)  
distribution (bin=0.25) of 55 ISO-detected sample galaxies with $z$$>$0.4 
in all of the three fields (the shaded region),
with the median IR luminosity of log($L_{\rm IR}/L_{\odot}$)=11.32,
which is similar to those of the local IRAS sample of V95 and Kim et al. (1995)
(the dotted-line for BGSs, and the dashed-line for WGSs, with the median values of
11.34 and 11.38, respectively). 
{\bf (b)} The distribution of the corresponding 38 galaxies with $z>0.4$ 
in the UDSR and UDSF fields (the solid line), with the median IR luminosity 
of log($L_{\rm IR}/L_{\odot}$)=11.27, 
and the distribution of the galaxies in the CFRS 3$^h$ field
(the dashed-line), with the median IR luminosity of log($L_{\rm IR}/L_{\odot}$)=11.55.
Although the ISOCAM survey observations in the UDSR and UDSF fields are 
three times deeper than in the CFRS 3$^h$ field,  
the total distribution of the galaxies in the 3 fields
is a reliable representation of 
the IR luminous galaxies in $z$$>$0.4 Universe.
}
\label{ISONfig}
\end{figure}

The IR luminosities (and deduced SFRs) have been calculated using the
procedure given in Elbaz et al. (2002) and are given in  Table 1 and 2. They
are based on mid-IR fluxes which show good correlations with radio and far-IR
measurements in the  local Universe (Elbaz et al. 2002).  In the distant
Universe, these estimates agree within a factor of 2 with those based on
H$\alpha$ luminosities (Flores et al. 2004a). Fig.~\ref{ISONfig}(a) shows the
distribution (the shaded region) of the inferred IR luminosity (8$-$1000
$\mu$m) of the 55 ISO/15$\mu$m-detected objects with $z>0.4$  (the called
``high-$z$" sample in the following parts of this paper) in the three fields
for VLT/FORS2 spectroscopic observation (see Sect.3 for redshifts) with a
median value of log($L_{\rm IR}/L_{\odot}$)=11.32.  Fig.~\ref{ISONfig}(b) 
shows the corresponding distribution of the 38 $z>0.4$ objects in the UDSR and
UDSF fields with a median value of log($L_{\rm IR}/L_{\odot}$)=11.27, and
the distribution of the objects in the CFRS 3$^h$ field with a median value of
11.55. The difference is simply related to the different flux limits
adopted in UDSR and UDSF fields on one side and on the CFRS 3$^h$ field, on
the other side. However, our high-$z$ sample exhibits IR luminosity
distribution very similar to local IRAS galaxies studied by V95 and Kim et al.
(1995),  in which the median log($L_{\rm IR}/L_{\odot}$)=11.34 for the Bright
Galaxies (BGSs) and 11.38 for the Warm Galaxies (WGSs) (in
Fig.~\ref{ISONfig}(a), the dotted-line for BGSs, and the dashed-line for
WGSs).  We believe that our sample can be used to probe the properties of
distant LIRGs over an IR luminosity range comparable to that of V95.

\subsection{Spectroscopic observations and data reduction}

Spectrophotometric observations of the 105 targets  were obtained during three
nights with the ESO 8m VLT using the FORS2 with R600, I600 at a resolution of
5\AA$~$and covering the possible wavelength range between 5000 and 9200\AA.
The slit width is 1.2{$^{\prime\prime}$} and the slit length is
10$^{\prime\prime}$. Spectra were extracted and wavelength-calibrated using
the IRAF{\footnote {IRAF is distributed by the National Optical Astronomical
Observatories, which is operated by the Association of Universities for
Research in Astronomy, Inc., under cooperative agreement with the National
Science Foundation.}} package. Flux calibration was done using 15 minute
exposures of 3  photometric standard stars per field. In addition, for one
field (CFRS), we have compared the spectrophotometry to the V and I
photometries and found a very good agreement. To ensure the reliability of
the data, all spectrum extractions as well as lines measurements were
performed by using the SPLOT program.

\begin{figure}
  \centering
  \includegraphics[bb=19 265 578 705,width=8.8cm,clip]{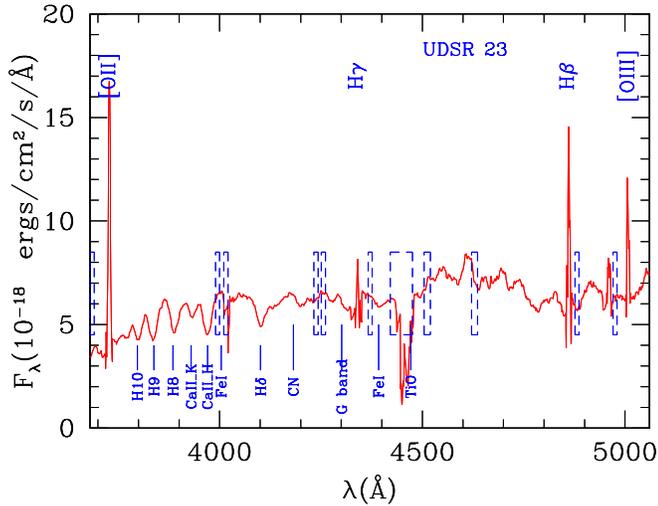}
\caption {Rest-frame spectrum of one of the sample galaxies, UDSR23.
It is a luminous infrared galaxy with log($L_{\rm IR}/L_{\odot}$)=11.38.
The continuum has been convolved except at the location of the 
emission lines (e.g. \ion{[O}{ii]} $\lambda$3727, H$\gamma$, H$\beta$ and
\ion {[O}{iii]} $\lambda$$\lambda$ 4959,5007). 
 The dashed boxes delimit the wavelength regions where
strong sky emission lines (e.g. $\ion{[O}{i]}$ 5577, 5891, 6300, 6364 \AA$~$
and OH 6834, 6871, 7914 \AA$~$ etc.) and
absorption lines (O$_2$ 6864, 7604 \AA$~$ etc.) are located. }
\label{spectrum}
\end{figure}

A rest-frame spectrum of one typical galaxy of our sample, UDSR23, is given in
Fig.~\ref{spectrum}.  The strong emission lines (e.g. $\ion{[O}{ii]}$
$\lambda$3727, H$\gamma$, H$\beta$, $\ion{[O}{iii]}$ $\lambda$5007) and the
obvious absorption lines are marked.  The continuum has been convolved except
at the locations of the marked emission lines using the softwares developed by
our group (Hammer et al. \cite{hammer01}; Gruel \cite{Gruel02}).  The adopted
convolution factors are 7 pixels and then 15 pixels.

{ 
\begin{table*} [b]
\scriptsize

\caption { Basic data of the sample galaxies in CFRS 3$^h$ field }  
\label{tab1}
\begin{tabular}{cccccccccrrr} \hline

Object  & RA   & DEC       & z & ${I_{\rm AB}}_{\rm phot}$ & ${I_{\rm AB}}_{\rm spec}$ & Aper &
M$_{\rm B,AB}$  & Type & f(S15) & log($\frac{L_{\rm IR}}{L_{\odot}}$) & SFR$_{\rm IR}$   \\ 
 Slit        & (J2000)& (J2000)   &   &      &    &     &            &
              & $\mu$Jy & & $M_{\odot}$\,yr$^{-1}$  \\ 
(1)    &    (2)    &  (3)    &  (4)   & (5)  & (6) &   (7)   & (8) &  (9)  & (10) & (11)  &   (12)     \\ \hline
03.1153  CFRS01 & 3 02 52.12 & +00 12 56.3 & 0.0000& 21.64 & 21.89 &1.26 &        ---      & Star  &   0.0 &  0.00           &      0.00       \\ 
03.1309  CFRS02 & 3 02 52.01 & +00 10 33.0 & 0.6172& 20.62 & 21.13 &1.60 &     $-$21.49    & EL    & 507.0 & 11.86$\pm$0.08  &    124.86  \\ 
03.1324  CFRS03 & 3 02 50.72 & +00 10 41.0 & 0.1748& 21.80 & 21.20 &0.58 &     $-$17.19    & EL    &   0.0 &  0.00           &      0.00 \\ 
03.1183  CFRS04 & 3 02 49.37 & +00 13 ~7.5 & 0.3870& 21.90 & 21.93 &1.03 &     $-$18.37    & Ear   & 419.0 & 11.26$\pm$0.10  &     31.06 \\ 
03.1188  CFRS05 & 3 02 48.85 & +00 12 30.2 & 9999  & 21.68 & 23.58 &5.75 &         ---     & ?     &   0.0 &   ---           &      ---  \\ 
03.1349  CFRS06 & 3 02 49.08 & +00 10 ~1.8 & 0.6169& 20.87 & 21.47 &1.74 &     $-$21.16    & EL    & 334.0 & 11.56$\pm$0.11  &     62.07 \\ 
03.1522  CFRS07 & 3 02 47.31 & +00 13 ~7.2 & 0.5870& 22.02 & 23.18 &2.91 &     $-$19.59    & ?     & 213.0 & 11.21$\pm$0.12  &     28.06  \\ 
03.0422  CFRS08 & 3 02 46.29 & +00 13 53.6 & 0.7154& 21.21 & 21.97 &2.01 &     $-$21.22    & EL    & 331.0 & 11.72$\pm$0.11  &     90.80 \\ 
03.0641  CFRS09 & 3 02 45.93 & +00 11 25.7 & 0.2613& 20.03 & 20.81 &2.05 &     $-$19.11    & EL    &   0.0 &  0.00           &      0.00 \\ 
03.0445  CFRS10 & 3 02 44.57 & +00 12 20.1 & 0.5276& 20.64 & 21.20 &1.70 &     $-$21.23    & EL    & 220.0 & 11.17$\pm$0.19  &     25.67 \\
03.1541  CFRS11 & 3 02 43.38 & +00 12 ~9.5 & 0.6895& 21.85 & 21.80 &1.06 &     $-$20.52    & EL    & 425.0 & 11.86$\pm$0.10  &    123.54 \\ 
03.0495  CFRS12 & 3 02 41.54 & +00 14 42.9 & 0.2614& 19.43 & 20.01 &2.05 &     $-$19.82    & EL    & 323.0 & 10.60$\pm$0.12  &      6.81 \\ 
03.0711  CFRS13 & 3 02 42.31 & +00 10 ~1.4 & 0.2615& 21.04 & 21.52 &1.55 &     $-$18.54    & EL    & 180.0 & 10.32$\pm$0.22  &      3.56 \\
03.0507  CFRS14 & 3 02 40.44 & +00 14 ~3.8 & 0.4652& 20.95 & 21.33 &1.42 &     $-$20.47    & EL    & 195.0 & 11.04$\pm$0.30  &     19.00 \\ 
03.1531  CFRS15 & 3 02 39.72 & +00 13 21.2 & 0.7163& 21.81 & 23.56 &4.96 &     $-$20.52    & EL    & 293.0 & 11.65$\pm$0.11  &     76.85 \\ 
03.0533  CFRS16 & 3 02 38.80 & +00 14 17.5 & 0.8274& 21.47 & 21.98 &1.60 &     $-$21.38    & EL    & 401.0 & 12.10$\pm$0.10  &    215.84  \\ 
03.0776  CFRS17 & 3 02 39.18 & +00 10 35.1 & 0.8837& 22.37 & 22.88 &1.60 &     $-$20.69    & EL    & 177.0 & 11.59$\pm$0.20  &     67.38 \\ 
03.0799  CFRS18 & 3 02 38.21 & +00 11 50.1 & 9999  & 24.04 & 25.47 &3.73 &         ---     & ?     &   0.0 &  ---            &       --- \\  
03.0828  CFRS19 & 3 02 37.44 & +00 11 17.4 & 0.3310& 22.44 & 23.29 &2.19 &     $-$17.83    & EL    & 305.0 & 10.88$\pm$0.12  &     12.95 \\ 
03.0569  CFRS20 & 3 02 35.63 & +00 14 12.0 & 0.1810& 21.33 & 22.32 &2.49 &     $-$16.73    & ?     & 176.0 &  9.88$\pm$0.23  &      1.32 \\ 
03.0578  CFRS21 & 3 02 35.19 & +00 14 10.1 & 0.2189& 20.79 & 22.27 &3.91 &     $-$18.62    & EL    & 167.0 & 10.07$\pm$0.23  &      2.04  \\ 
03.0589  CFRS22 & 3 02 34.41 & +00 13 31.9 & 0.7172& 22.18 & 22.53 &1.38 &     $-$20.25    & EL    &   0.0 &  0.00           &     0.00  \\ 
03.5004  CFRS23 & 3 02 34.82 & +00 12 42.9 & 0.0875& 17.74 & 19.63 &5.68  &       ---      & EL    &2089.0 & 10.29$\pm$0.08  &      3.35 \\
03.0916  CFRS24 & 3 02 33.71 & +00 11 37.7 & 1.0300& 21.31 & 22.05 &1.98 &        ---      & QSO   & 219.0 & 12.04$\pm$0.12  &      ---  \\
03.0932  CFRS25 & 3 02 33.14 & +00 11 ~5.2 & 0.6478& 21.37 & 22.21 &2.17 &     $-$20.73    & EL    & 529.0 & 11.94$\pm$0.09  &    150.83 \\
03.0006  CFRS26 & 3 02 31.53 & +00 14 ~3.3 & 0.8836& 23.29 & 23.94 &1.82 &     $-$19.77    & EL    & 183.0 & 11.61$\pm$0.23  &     69.65 \\
03.0015  CFRS27 & 3 02 30.73 & +00 12 58.6 & 0.1980& 21.68 & 21.64 &0.96 &     $-$16.60    & Ear   &   0.0 &  0.00           &      0.00 \\
03.0138  CFRS28 & 3 02 30.82 & +00 11 ~1.8 & 0.4797& 20.09 & 20.46 &1.41 &     $-$20.87    & Ear   &   0.0 &  0.00           &      0.00  \\
03.0035  CFRS29 & 3 02 29.41 & +00 12 59.8 & 0.8804& 21.16 & 21.99 &2.15 &     $-$21.88    & EL    & 318.0 & 12.03$\pm$0.12  &    182.36 \\
03.0149  CFRS30 & 3 02 29.51 & +00 ~9 17.2 & 0.2510& 20.74 & 22.29 &4.17 &     $-$18.92    & EL    &   0.0 &  0.00           &      0.00 \\
03.0062  CFRS31 & 3 02 27.20 & +00 13 56.6 & 0.8272& 20.96 & 21.79 &2.15 &     $-$21.85    & EL    & 228.0 & 11.67$\pm$0.15 &     80.51 \\
03.0174  CFRS32 & 3 02 27.53 & +00 10 57.2 & 0.5251& 22.31 & 23.06 &2.00 &     $-$19.11    & EL    & 212.0 & 11.16$\pm$0.26 &     24.66 \\
03.0186  CFRS33 & 3 02 26.86 & +00 10 55.1 & 0.5254& 22.01 & 22.88 &2.23 &     $-$19.32    & EL    & 185.0 & 11.08$\pm$0.25 &     20.58 \\
03.0085  CFRS34 & 3 02 25.24 & +00 13 24.8 & 0.6089& 22.00 & 22.29 &1.31 &     $-$19.96    & EL    & 203.0 & 11.21$\pm$0.20 &     27.66 \\ \hline  
 \end{tabular} 

{ \scriptsize  Note: The magnitude of CFRS23 is $R_{\rm AB}$ and is not $I_{\rm AB}$. }

\end{table*} 
}

{
\begin{table*} 
\centering
 \tiny
\caption { Basic data of the sample galaxies in UDSR and UDSF fields} 
 \label{tab2}
  \begin{tabular}{lccccccccrrr} \hline
Object   & RA   & DEC       & z & ${R_{\rm AB}}_{\rm phot}$   &${R_{\rm AB}}_{\rm spec}$   &  Aper  &  M$_{\rm B,AB}$ & Type &
f(S15) & log($\frac{L_{\rm IR}}{L_{\odot}}$) & SFR$_{\rm IR}$   \\ 
 Slit         & (J2000)& (J2000)   &   &         &   &        &      &  &
$\mu$Jy &   & $M_{\odot}$\,yr$^{-1}$ \\
(1)    &    (2)    &  (3)    &  (4)   & (5)  & (6) &   (7)   & (8) &  (9)  & (10) & (11)  &   (12)     \\\hline
 UDSR01  &  3 15 02.4 & $-55$ 21 25 & 0.1025 & 20.94 & 21.72 &  2.05  &     $-$17.03  & EL   &    91.2   &   8.91$\pm$0.20  &    0.14      \\ 
 UDSR02  &  3 14 54.6 & $-55$ 22 17 & 0.9334 & 23.16 & 23.72 &  1.67  &     $-$20.72  & EL   &     0.0   &   0.00           &    0.00    \\ 
 UDSR03  &  3 14 51.6 & $-55$ 22 27 & 9999   & 23.28 & 23.77 &  1.57  &         ---   & ?    &     0.0   &    ---           &    ---    \\ 
 UDSR04  &  3 14 55.9 & $-55$ 21 34 & 0.4501 & 22.97 & 23.21 &  1.25  &     $-$19.42  & EL   &   231.5   &  10.96$\pm$0.18  &   15.68   \\ 
 UDSR05  &  3 15 07.1 & $-55$ 19 45 & 1.0657 & 23.22 & 23.74 &  1.61  &     $-$22.18  & EL   &   249.6   &  12.03$\pm$0.24  &  183.10    \\ 
 UDSR06  &  3 15 05.1 & $-55$ 19 51 & 0.1836 & 19.88 & 20.32 &  1.50  &     $-$19.30  & EL   &   345.5   &  10.08$\pm$0.12  &    2.07   \\ 
 UDSR07  &  3 14 57.3 & $-55$ 20 31 & 0.5507 & 23.91 & 24.36 &  1.51  &     $-$19.27  & ?    &   251.9   &  11.10$\pm$0.19  &   21.85   \\ 
 UDSR08  &  3 14 55.2 & $-55$ 20 31 & 0.7291 & 21.60 & 22.18 &  1.71  &     $-$21.52  & EL   &   235.7   &  11.32$\pm$0.21  &   35.66   \\ 
 UDSR09  &  3 14 56.1 & $-55$ 20 08 & 0.3884 & 19.53 & 19.95 &  1.47  &     $-$21.75  & EL   &   608.9   &  11.30$\pm$0.08  &   34.21   \\ 
 UDSR10  &  3 14 43.9 & $-55$ 21 35 & 0.6798 & 20.84 & 21.48 &  1.80  &     $-$22.29  & EL   &   495.2   &  11.75$\pm$0.13  &   96.92   \\ 
 UDSR11  &  3 14 45.7 & $-55$ 20 53 & 0.1660 & 19.81 & 20.38 &  1.69  &     $-$19.20  & EL   &   252.1   &   9.82$\pm$0.13  &    1.15   \\ 
 UDSR12  &  3 14 37.5 & $-55$ 21 44 & 0.1654 & 20.76 & 21.15 &  1.43  &     $-$18.23  & EL   &   226.7   &   9.77$\pm$0.15  &    1.01    \\ 
 UDSR13  &  3 14 35.9 & $-55$ 21 35 & 0.1666 & 20.78 & 21.12 &  1.37  &     $-$18.15  & EL   &    53.0   &   9.14$\pm$0.23  &    0.24    \\ 
 UDSR14  &  3 14 43.3 & $-55$ 20 11 & 0.8150 & 21.56 & 22.29 &  1.96  &     $-$22.22  & EL   &   200.1   &  11.34$\pm$0.24  &   38.00   \\ 
 UDSR15  &  3 14 36.3 & $-55$ 21 02 & 0.4949 & 21.51 & 22.19 &  1.87  &     $-$20.61  & EL   &   235.6   &  11.03$\pm$0.19  &   18.40   \\ 
 UDSR16  &  3 14 33.1 & $-55$ 21 01 & 0.1662 & 20.86 & 21.15 &  1.31  &     $-$18.29  & EL   &   180.4   &   9.67$\pm$0.16  &    0.82   \\ 
 UDSR17  &  3 14 41.3 & $-55$ 19 24 & 0.9680 & 22.51 & 23.05 &  1.64  &     $-$22.25  & EL   &   386.2   &  12.10$\pm$0.19  &  219.01   \\ 
 UDSR18a &  3 14 53.5 & $-55$ 17 26 & 9999   & 23.33 & 23.24 &  0.92  &         ---   & ?    &     0.0   &   ---            &     ---   \\ 
 UDSR18b &  3 14 53.4 & $-55$ 17 24 & 9999   & 24.43 & 23.74 &  0.53  &         ---   & ?    &   247.1   &   ---            &     ---    \\
 UDSR19a &  3 14 39.5 & $-55$ 19 03 & 0.0000 & 22.53 & 22.48 &  0.95  &         ---   & Star &   174.7   &   0.00           &    0.00    \\ 
 UDSR19b &  3 14 39.3 & $-55$ 19 01 & 0.0000 & 22.66 & 22.98 &  1.34  &         ---   & Star &   174.7   &   0.00           &    0.00   \\ 
 UDSR20  &  3 14 41.1 & $-55$ 18 40 & 0.7660 & 21.79 & 23.12 &  3.40  &     $-$21.87  & EL   &   214.4   &  11.32$\pm$0.23  &   35.81   \\ 
 UDSR21  &  3 14 43.6 & $-55$ 17 50 & 0.4655 & 22.54 & 23.17 &  1.79  &     $-$19.52  & EL   &    56.8   &  10.24$\pm$0.26  &    3.01   \\ 
 UDSR22  &  3 14 48.2 & $-55$ 16 57 & 0.4970 & 21.92 & 23.15 &  3.10  &         ---   & ?    &     0.0   &   0.00           &    0.00   \\ 
 UDSR23  &  3 14 32.1 & $-55$ 19 02 & 0.7094 & 21.12 & 21.45 &  1.36  &     $-$21.70  & EL   &   271.1   &  11.38$\pm$0.20  &   40.96   \\ 
 UDSR24  &  3 14 33.7 & $-55$ 18 40 & 0.3136 & ---   & 20.46 &  ---   &         ---   & Ear  &   164.0   &  10.36$\pm$0.18  &    3.96   \\ 
 UDSR25a &  3 14 24.8 & $-55$ 19 35 & 0.5816 & 20.48 & 20.61 &  1.13  &     $-$22.08  & EL   &   702.5   &  11.85$\pm$0.09  &  122.08   \\ 
 UDSR25b &  3 14 24.8 & $-55$ 19 36 &  9999  & 21.11 &  ---  &  ---   &         ---   & ?    &     0.0   &    ---           &    ---     \\ 
 UDSR26  &  3 14 36.7 & $-55$ 17 32 & 0.3841 & 21.08 & 21.68 &  1.74  &     $-$20.27  & EL   &    82.5   &  10.27$\pm$0.23  &    3.17    \\ 
 UDSR27  &  3 14 29.6 & $-55$ 18 22 & 0.6429 & 22.19 & 22.18 &  0.99  &     $-$20.61  & EL   &   128.9   &  10.80$\pm$0.23  &   10.93   \\ 
 UDSR28  &  3 14 20.7 & $-55$ 19 20 & 0.5210 & 20.65 & 21.01 &  1.39  &     $-$21.76  & Ear  &   231.5   &  11.04$\pm$0.18  &   18.65    \\ 
 UDSR29  &  3 14 27.4 & $-55$ 18 06 & 9999   & 23.52 & 23.43 &  0.92  &         ---   & ?    &     0.0   &    ---           &     ---    \\ 
 UDSR30  &  3 14 26.7 & $-55$ 17 51 & 0.9590 & 22.69 & 23.08 &  1.43  &     $-$21.45  & EL   &   408.7   &  12.12$\pm$0.18  &  225.95    \\ 
 UDSR31  &  3 14 25.6 & $-55$ 17 51 & 0.9530 & 22.84 & 23.17 &  1.36  &     $-$21.81  & ?    &   274.5   &  11.80$\pm$0.21  &  109.40     \\ 
 UDSR32  &  3 14 30.3 & $-55$ 16 53 & 0.5841 & 22.33 & 22.96 &  1.79  &     $-$19.98  & EL   &   137.3   &  10.77$\pm$0.23  &   10.13   \\\hline 
 
 UDSF01  &  3 13 27.6 & $-55$ 05 53 & 0.4656 & 20.41 & 21.22 &  2.11 &     $-$21.61   & EL   &    86.6  &   10.45$\pm$0.11   &      4.82   \\ 
 UDSF02  &  3 13 23.3 & $-55$ 05 15 & 0.7781 & 22.57 & 24.00 &  3.73 &     $-$21.43   & EL   &    47.0  &   10.43$\pm$0.12   &      4.59   \\ 
 UDSF03  &  3 13 33.6 & $-55$ 04 27 & 0.5532 & 21.54 & 22.65 &  2.78 &     $-$20.86   & EL   &     0.0  &$<$10.81            &$<$11.14   \\ 
 UDSF04  &  3 13 28.5 & $-55$ 04 45 & 0.9620 & 22.99 & 23.44 &  1.51 &     $-$21.54   & EL   &   246.8  &   11.75$\pm$0.12   &   97.17   \\ 
 UDSF05  &  3 13 30.4 & $-55$ 04 17 & 0.1658 & 19.85 & 20.46 &  1.75 &     $-$19.12   & Ear  &     0.0  & $<$9.59            & $<$0.69   \\ 
 UDSF06  &  3 13 30.2 & $-55$ 04 04 & 0.6928 & 21.91 & 22.31 &  1.45 &     $-$21.52   & EL   &   150.0  &   10.97$\pm$0.33  &   15.97   \\ 
 UDSF07  &  3 13 17.3 & $-55$ 05 16 & 0.7014 & 22.37 & 22.95 &  1.71 &     $-$20.70   & EL   &   154.3  &   10.99$\pm$0.12  &   16.77   \\ 
 UDSF08  &  3 13 14.2 & $-55$ 05 37 & 0.7075 & 22.30 & 22.96 &  1.84 &     $-$21.03   & EL   &   137.7  &   10.91$\pm$0.12  &   14.01   \\ 
 UDSF09  &  3 13 18.0 & $-55$ 04 57 & 0.2273 & 19.99 & 20.93 &  2.38 &     $-$19.71   & Ear  &    68.9  &    9.58$\pm$0.10  &    0.69   \\ 
 UDSF10  &  3 13 12.6 & $-55$ 05 11 & 9999   & 24.04 & 24.31 &  1.28 &          ---   & ?    &     0.0  &    ---            &    ---    \\ 
 UDSF11  &  3 13 23.3 & $-55$ 03 33 & 0.0937 & 19.36 & 19.93 &  1.69 &     $-$18.39   & EL   &   155.4  &    9.06$\pm$0.09  &    0.23   \\ 
 UDSF12  &  3 13 07.7 & $-55$ 05 26 & 0.7388 & 22.45 & 23.13 &  1.87 &     $-$21.39   & EL   &   331.1  &   11.56$\pm$0.13  &   62.83   \\ 
 UDSF13  &  3 13 14.2 & $-55$ 04 21 & 0.7605 & 23.54 & 24.37 &  2.15 &     $-$20.44   & EL   &   906.6  &   12.38$\pm$0.11  & 408.45   \\ 
 UDSF14  &  3 13 09.7 & $-55$ 04 33 & 0.8190 & 23.08 & 23.92 &  2.17 &     $-$21.08   & EL   &    64.1  &   10.65$\pm$0.13  &   7.70    \\ 
 UDSF15  &  3 13 10.7 & $-55$ 04 06 & 0.2302 & 22.60 & 23.55 &  2.40 &     $-$17.57   & EL   &     0.0   &$<$9.95           &$<$1.49    \\ 
 UDSF16  &  3 13 08.0 & $-55$ 04 18 & 0.4548 & 20.39 & 21.11 &  1.94 &     $-$21.47   & EL   &   137.6  &   10.68$\pm$0.11  &     8.16    \\ 
 UDSF17  &  3 13 08.6 & $-55$ 03 57 & 0.8100 & 21.45 & 22.16 &  1.92 &     $-$22.42   & EL   &   256.9  &   11.52$\pm$0.14  &  56.97    \\ 
 UDSF18  &  3 13 16.5 & $-55$ 02 27 & 0.4620 & 21.42 & 22.08 &  1.84 &     $-$20.60   & EL   &   170.0  &   10.81$\pm$0.10  &  11.03    \\ 
 UDSF19  &  3 13 09.8 & $-55$ 03 08 & 0.5476 & 20.68 & 21.20 &  1.61 &     $-$21.94   & EL   &   611.3  &   11.67$\pm$0.11  &  80.86    \\ 
 UDSF20  &  3 13 19.0 & $-55$ 01 42 & 0.8424 & 22.30 & 23.25 &  2.40 &         ---    & EL   &   117.2  &   11.04$\pm$0.14  &  18.84     \\ 
 UDSF21  &  3 13 04.5 & $-55$ 03 12 & 0.6980 & 21.78 & 22.41 &  1.79 &     $-$21.35   & EL   &     0.0  &$<$10.97           &$<$16.20     \\ 
 UDSF22  &  3 12 51.7 & $-55$ 04 39 & 0.2298 & 20.48 & 20.88 &  1.45 &     $-$19.38   & EL   &   134.1  &    9.89$\pm$0.09  &     1.38     \\
 UDSF23  &  3 13 05.0 & $-55$ 02 36 & 1.0404 & 22.76 & 23.26 &  1.58 &     $-$21.83   & ?    &     0.0  &$<$11.58           &$<$65.47     \\ 
 UDSF24  &  3 13 04.8 & $-55$ 02 24 & 1.1586 & 21.92 & 23.13 &  3.05 &     $-$23.31   & EL   &   159.7  &   11.91$\pm$0.16  & 138.87      \\ 
 UDSF25  &  3 13 07.3 & $-55$ 01 56 & 0.8094 & 23.32 & 24.94 &  4.45 &     $-$20.42   & EL   &    94.2  &   10.86$\pm$0.13  &   12.52     \\ 
 UDSF26a &  3 12 58.0 & $-$55 02 53 & 0.7027 & 23.07 & 24.63 &  4.21 &     $-$21.60   & EL   &     0.0  &$<$10.98           &$<$16.31     \\
 UDSF26b &  3 12 57.5 & $-55$ 02 48 & 0.7023 & 21.16 & 21.83 &  1.85 &     $-$20.62   & EL   &   163.3  &   11.03$\pm$0.13  &  18.38     \\ 
 UDSF27a &  3 12 58.4 & $-$55 02 34 & 9999   & 23.43 & 23.98 &  1.66 &         ---    &  ?   &   108.1  &   ---             &    ---     \\
 UDSF27b &  3 12 58.2 & $-55$ 02 32 & 9999   & 23.72 & 24.33 &  1.75 &         ---    & ?    &     0.0  &   ---             &    ---      \\ 
 UDSF28  &  3 12 51.8 & $-55$ 02 57 & 0.6612 & 21.31 & 22.25 &  2.38 &     $-$21.84   & EL   &   321.4  &   11.40$\pm$0.12  &  42.96     \\ 
 UDSF29a &  3 12 50.5 & $-55$ 03 ~1 & 9999   & 24.18 & ---   &   --- &         ---    &  ?   &     0.0  &   ---             &   ---      \\
 UDSF29b &  3 12 50.2 & $-55$ 02 59 & 0.6619 & 22.64 & 23.37 &  1.96 &     $-$20.45   & EL   &   353.9  &   11.47$\pm$0.12  &  51.34     \\ 
 UDSF30  &  3 12 44.9 & $-55$ 03 27 & 9999   & 23.31 & 22.79 &  0.62 &         ---    & ?    &   114.6  &   ---             &   ---      \\ 
 UDSF31  &  3 12 44.0 & $-55$ 03 21 & 0.6868 & 22.39 & 22.93 &  1.64 &     $-$20.57   & EL   &   192.8  &   11.10$\pm$0.13  &  21.71   \\ 
 UDSF32  &  3 12 53.3 & $-55$ 01 43 & 0.7268 & 23.16 & 23.78 &  1.77 &     $-$19.73   & EL   &   226.3  &   11.27$\pm$0.13  &  32.28   \\ 
 UDSF33  &  3 12 53.9 & $-55$ 01 28 & 9999   & 21.07 & 22.17 &  2.75 &         ---    & ?    &   295.7  &   ---             &    ---   \\   \hline 
                                                                                                                                          
 \end{tabular} 
\end{table*} 
}

\section{Redshift distributions and spectral types}

\begin{figure} 
   \centering
   \includegraphics[bb=70 320 412 658,width=8.8cm,clip]{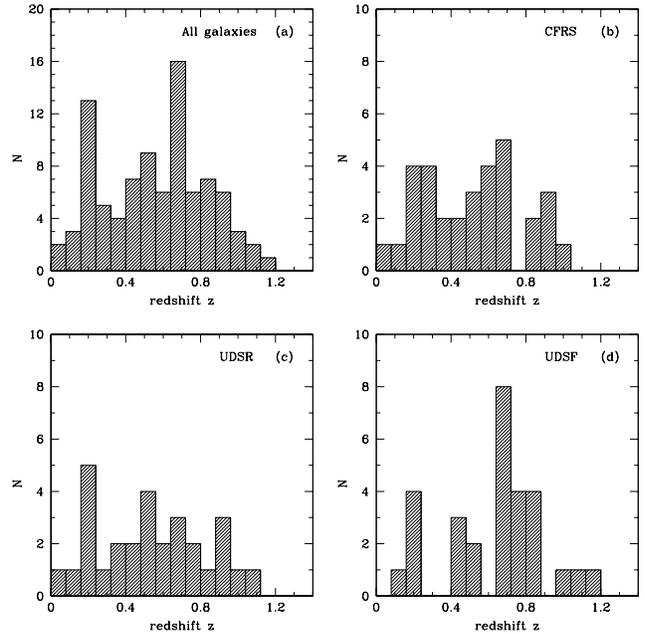}
   \caption{ Redshift distributions (bin=0.08) of the sample
galaxies in the combined and the individual three fields.}  
    \label{zNfig}
     \end{figure}

Redshifts are identified by  using the emission and/or absorption lines.
Col.(4) of Table~\ref{tab1} and \ref{tab2} gives the $z$ values of the
objects.  Redshift distributions of the combined and the individual three
fields are shown in Fig.~\ref{zNfig}. The corresponding median redshifts are
$z_{\rm med}$=0.587 in the combined sample, $z_{\rm med}$=0.525 in the CFRS, 
$z_{\rm med}$=0.521 in the UDSR and $z_{\rm med}$=0.698 in the UDSF. The
redshift peak around z=0.70 in the UDSF field (six galaxies) shows a velocity
dispersion of $\sim$1390 km$~$s$^{-1}$, a typical value for a galactic
cluster. The corresponding six objects are UDSF06, 07, 08, 21, 26a and 26b. In
the UDSR field, four objects (UDSR11, 12, 13 and 16) show a redshift peak
around z=0.166, which may correspond to a velocity dispersion of $\sim$129
km$~$s$^{-1}$, a typical value for a galactic group.

The redshift distributions are consistent with  the results in some other
ISOCAM survey fields, e.g., the $z_{\rm med}$=0.7 in the CFRS 14$^h$ field
(Flores et al. 1999)  and 0.585 in the HDF-N (Aussel et al. 1999).
Franceschini et al. (2003) found a  peak at $z\sim 0.6$ for their 21 objects
in the Hubble Deep Field South (HDF-S) field,  which they suggested to be a
cluster  or a large galaxy concentration.  The similarities between the
IR luminosities as redshift distribution of our  sample to those of other
studies lead us to assume that our sample can be used to test the properties
of distant LIRGs. About 81\% (75/92) of the redshift-identified galaxies show
obvious and strong emission line (EL)  (see Col.(9) of Tables~\ref{tab1} and
\ref{tab2}). The corresponding EL galaxies fraction of ISO-detected objects 
is $\sim$85\%. Table~\ref{tab3} summarizes the redshift identifications and
spectral types of the galaxies in our sample. 

{ 
\begin{table*} 
{ \scriptsize
\caption { Redshift-identification and spectral types of the
galaxies in the three fields. 
``ELGs" means ``Emission Line Galaxies", 
``ETGs" means ``Early-Type Galaxies",  
``z-poor" means the redshift can be gotten 
even the Type is not clear from the spectrum, }  

\label{tab3}

 \begin{tabular}{cc|cccccccc|ccccc} \hline
Fields & Objects & \multicolumn{3}{c} {ELGs}                                            &  ETGs & Stars &  QSO & z-     & z- & \multicolumn{2}{c} {Redshifts}                       & Redshift  & median     \\ \cline{3-5}\cline{11-12}
       &         & \multicolumn{2}{c}{ISO}    &\multicolumn{1}{c}{non-ISO}              &       &       &      & poor   & unknown & \multicolumn{1}{c}{ISO} &\multicolumn{1}{c}{non-ISO}&  identified  &  $z$          \\ \cline{3-4}
       &         & \multicolumn{1}{c}{$z$$>$0.4} & \multicolumn{1}{c}{$z$$\leq$0.4}     &       &       &      &        &                & & &   &                \\ \hline
CFRS   & 34      & \multicolumn{1}{c}{16} & \multicolumn{1}{c}{5} &\multicolumn{1}{c}{4} & 3    &  1    &  1   & 2      &   2          & \multicolumn{1}{c}{25} &\multicolumn{1}{c}{7} & 94.12 \% &  0.525      \\ 
UDSR   & 35      & \multicolumn{1}{c}{14} & \multicolumn{1}{c}{8} &\multicolumn{1}{c}{1} & 2    &  2    &  0   & 3      &   5          & \multicolumn{1}{c}{28} &\multicolumn{1}{c}{2} &  85.71 \% &  0.521     \\  
UDSF   & 36      & \multicolumn{1}{c}{21} & \multicolumn{1}{c}{2} &\multicolumn{1}{c}{4} & 2    &  0    &  0   & 1      &   6          & \multicolumn{1}{c}{24} &\multicolumn{1}{c}{6} &  83.33 \% & 0.698        \\ \hline
Three fields& 105& \multicolumn{1}{c}{51} & \multicolumn{1}{c}{15} & \multicolumn{1}{c}{9} & 7  & 3     &  1   &  6     &   13          & \multicolumn{1}{c}{77} &\multicolumn{1}{c}{15}  &   87.62 \%  &  0.587    \\
\hline
 \end{tabular} 
}
\end{table*} 
}

\section{Flux measurements, extinction and SFRs}
\subsection{Flux measurements}

The fluxes of the emission lines are measured using the SPLOT package.   The
stellar absorption under the Balmer lines is estimated from the synthesized
stellar spectra obtained using the stellar spectra of Jacoby et al.
(\cite{Jacoby84}).  To do so, we use the spectra of four stellar types (e.g.
A, B, F and G types)  to synthesize the ``galactic" continuum and absorption
lines. Then, the ``pure" emission Balmer lines are obtained through reducing
the underlying stellar absorption. The corresponding error budget of
emission line flux is deduced by a quadratic addition of three independent
errors:  the first one is related to the use of stellar templates to fit
stellar absorption lines and continuum; the second one   comes from the
differences among independent measurements performed by Y.C. Liang, H. Flores
and F. Hammer; the third one is from the Poisson noises from both sky and
objects, and it actually dominates the error budget. The flux
measurements of emission lines and their percent errors are given in 
Table~\ref{Fluxtab} for high-$z$ galaxies, and in Table~\ref{Fluxtablow} for
low-$z$ galaxies. Three low-$z$ galaxies are also given in Table~\ref{Fluxtab}
for their H$\gamma$ fluxes. To obtain reliable global fluxes of emission lines
of the galaxies,  we should notice that the 1.2$\arcsec$  slit of VLT
observations does not always contain the whole galaxy. Thus, the fluxes of the
emission lines are corrected by an aperture factor derived by comparing the
photometric magnitudes to the spectral magnitudes at $I_{\rm AB}$ (for CFRS
field) or $R_{\rm AB}$ (for UDSR and UDSF fields) bands.  The aperture
correction factors are given in Col.(7) of Table~\ref{tab1} and \ref{tab2}.
 
\begin{table*} 
{ \begin{center} 
\scriptsize
 \caption [] {Measured emission line fluxes (F$_\lambda$) in unit of 10$^{-17}$ (ergs cm$^{-2}$
   s$^{-1}$ ) and the errors in percent for the high-$z$ EL galaxies.
  ``9997" means
the line is blended with strong sky line, ``9998" means there is no
corresponding emission line detected at the line position, and
``9999" means the line is shifted outside of the rest-frame wavelength
range. ``o5" means \ion{[O}{iii]}$_{5007}$, and ``o4" means \ion{[O}{iii]}$_{4959}$. } 
  \label{Fluxtab}

\begin{tabular}{lcrrrrrrrrrr} \hline
 Slit   &  z&     \ion{[O}{ii]}$_{3727}$  &  e$_{\ion{[O}{ii]}}$\% &  H$\gamma$ & e$_{H\gamma}$\% & 
H$\beta$ &  e$_{H\beta}$\% & \ion{[O}{iii]}$_{5007}$  &  e$_{o5}$\% &  \ion{[O}{iii]}$_{4959}$  &  e$_{o4}$\% \\ [1mm]  
  (1) & (2) & (3)$~$ & (4)$~$ & (5) & (6)$~$ & (7) & (8)$~$ & (9)$~$ & (10) & (11)$~$ & (12) \\  \hline

 CFRS & & & & & & &  & & & &\\
     2  &    0.6172  & 14.66   &  0.02   & 1.99    &  0.12  &  6.30  &   0.10  & 9.49   & 0.06 & 9998 & 9998     \\
     6  &    0.6169  &  6.55   &  0.02   & 1.63    &  0.10  &  5.04  &   0.10  & 8.49   &0.06  & 2.28 &  0.17    \\
     8  &    0.7154  &  4.28   &   0.04  &  1.38   &  0.13  &  4.37  &   0.10  &  9999  & 9999 & 9999 &  9999    \\
    10  &    0.5276  &  7.57   &   0.05  &  1.68   &  0.08  &  4.29  &   0.07  &  0.90  & 0.16 & 9998 & 9998    \\
    11  &    0.6895  &  5.10   &   0.09  &  1.05   &  0.20  &  3.92  &   0.08  &   1.18 & 0.25 & 9998 & 9998    \\
    12  &    0.2614  &  9999   &   9999  &  3.78   &  0.13  &  10.09 &  0.05   &  30.90 &  0.02 & 10.97  &   0.04 \\
    14  &    0.4652  &  16.81  &   0.02  &  2.09  &    0.11 &  6.87  &   0.05  &  6.55  & 0.10 & 2.55 &   0.17    \\
    15  &    0.7163  &   2.86  &   0.08  &  2.06  &    0.16 &  9997  & 9997    &  9998  & 9998 & 9998 & 9998    \\
    16  &    0.8274  &   11.42 &   0.04  &  1.64  &    0.15 &  9999  & 9999    &  9999  & 9999 & 9999 &  9999    \\
    17  &    0.8837  &   7.14  &   0.04  &  0.58  &    0.19 &  2.05  &   0.11  &  0.58  & 0.40 & 9998 & 9998    \\
    19  &    0.3310  &   9999  &   9999  & 0.29   &   0.16  & 0.77  &   0.08  &  1.44  & 0.10 & 0.43  &   0.32 \\
    22  &    0.7172  &   4.99  &   0.05  &  0.63  &   0.28  &  9997  & 9997    &  9999   & 9999 & 9999 & 9999    \\
    25  &    0.6478  &   6.34  &  0.04   &  1.14  &    0.17 &  2.87 &  0.15   &  2.85  & 0.11 & 9998 & 9998    \\
    26  &    0.8836  &   2.05  &   0.15  &   9998  &    9998 &  9999  & 9999    &  9999   & 9999 & 9999 &  9999    \\
    29  &    0.8804  &   2.38  &   0.16  &   9998  &    9998 &  9999  & 9999    &  9999   & 9999 & 9999 &  9999    \\
    31  &    0.8272  &   5.61  &   0.05  &   1.46  &    0.15 &  9999  & 9999    &  9999   & 9999 & 9999 & 9999    \\
    32  &    0.5251  &   1.13  &  0.13   &   9998  &   9998  &  0.31  &  0.10   &  0.32   & 0.16 & 9999 & 9999    \\
    33  &    0.5254  &   2.80  &  0.04   &  0.31   &   0.50  &  0.98 &   0.10  &   0.56 & 0.16 & 9998 & 9998    \\
    34  &    0.6089  &   9.97 &  0.04   &   0.76   &   0.15  &  9997  &  9997   &  2.12  & 0.22 & 9998 & 9998    \\ \hline

  UDSR & & & & & & & & & & & \\
     2  &    0.9334  &   5.39   &   0.05  &   1.01  &   0.30 &   9999 & 9999&  9999  &9999 & 9999 &9999    \\
     4  &    0.4501 &     3.71  &   0.08   &  9998   &   9998 &    0.58  &   0.15 &  1.65  & 0.13 & 0.51 & 0.23    \\
     5  &    1.0657 &     5.92  &   0.06  &   9999   &   9999 &   9999&  9999  &9999 & 9999 &9999 &9999    \\
     8  &    0.7291 &    23.27  &   0.02  &   4.07  &   0.07 &   9.52   &  0.06 &   10.88 & 0.05 &  5.03   &    0.07    \\
     9  &    0.3884 &    5.83   &   0.08  &   9998   &   9998  &   0.47   &  0.10 &    9.34 & 0.05 &  3.10  &  0.15    \\
    10  &    0.6798 &    11.13  &   0.06  &   2.47  &   0.15 &   7.02    &  0.09 &   5.24  & 0.13  &   1.830  &   0.15    \\
    14  &    0.8150 &    18.13  &   0.04 &    3.32  &   0.18  &   9.29   & 0.50   &  8.93  & 0.10  &   3.01 & 0.15    \\
    15  &    0.4949 &     9999   &   9999  &  1.20   &   0.10  &  4.02    &  0.15  &   3.92 &0.06  &   1.32  &   0.12    \\
    17 &     0.9680 &    2.28   &   0.25 & 9998 & 9998 &  9999  &9999 & 9999 & 9999 & 9999 & 9999\\
    20 &     0.7660 &    16.57  &   0.02 &    2.33  &   0.09 &   6.48   &   0.06 &   9.81 & 0.08  &9998  &9998    \\
    21 &     0.4655 &     2.03  &   0.13 &     9998 &    9998 &    0.45  &   0.40 &  4.75   &0.10 & 9998&  9998    \\
    23 &     0.7094 &    13.35  &   0.02 &    2.22 &    0.11 &   7.01   &   0.07 &  4.63   & 0.10 &  1.573  &   0.15    \\
    26 &     0.3841 &     3.61  &   0.10 &     9998 &    9998 &    1.25 &    0.20 &   1.04  & 0.16 & 9998 & 9998    \\
    27 &     0.6429 &     4.13  &   0.08 &    9998 &     9998 &    9997  &    9997 &    9998 & 9998 & 9998  &9998    \\
    30 &     0.9590 &     2.85  &   0.18 &    9998 &     9998 & 9999 & 9999 & 9999  &9999 &9999 &9999    \\
    32 &     0.5841 &     8.45  &   0.05 &  0.89    &   0.25 &  2.64    &   0.10  &   4.47 & 0.09 & 9998  &9998    \\  \hline

  UDSF & &  & &  & & & & & & & \\
     1 &     0.4656 &    30.79  &   0.02 &   3.74   &   0.08 &  9.42   &    0.04  & 13.98  & 0.11 &    5.11 &   0.12    \\
     2 &     0.7781 &    1.82   &   0.07 &  9999    &   9999 &  0.54   &    0.40  &  0.36  & 0.40 & 9998 & 9998    \\
     3 &     0.5532 &    2.66   &   0.09 &  9999    &   9999 &  1.90   &    0.13  &   1.45 & 0.30 & 9998 & 9998    \\
     4 &     0.9620 &    2.985  &   0.09 &    9998 & 9998 & 9999 & 9999 & 9999  &9999&  9999 & 9999   \\
     6 &     0.6928 &     4.70  &   0.11 &    9998&  9998 & 9998 & 9998 & 9999 & 9999 & 9998 &9998   \\
     7 &     0.7014 &     3.79  &   0.08 &    0.88  &   0.23 &    2.42  &   0.16  &   1.19 &  0.30 & 9998 & 9998    \\
     8  &    0.7075 &     2.15  & 0.08 & 9998&9998 & 9998&9998 & 9998& 9998&9998 & 9998\\  
    12 &     0.7388 &     4.44  &   0.06  &   9998   &   9998 & 1.84 & 0.11 & 9998 & 9998 &9998 & 9998    \\
    13 &     0.7605 &     5.62  &   0.08  &   0.51  &   0.30 &   1.58   &   0.40 &   21.55  & 0.05 &  6.02    &   0.10    \\
    14 &     0.8190 &     1.39  &   0.50  &   9998   &   9998  & 9998 & 9998 & 9998 & 9998 & 9998 & 9998 \\
    15 &     0.2302 &     9999  &   9999  &   0.74   &  0.10   &  1.87  &   0.10  &   3.92  &   0.05   &  1.49  &   0.11 \\
    16 &     0.4548 &     3.79  &   0.07  &   1.15  &   0.24 &   3.88    &  0.09  &  2.35  & 0.30 & 9998 & 9998    \\
    17 &     0.8100 &    10.57  &   0.06  &   1.45  &   0.33  &   4.09  &   0.18  &   1.56  &0.17 & 9998 & 9998    \\
    18  &    0.4620 &     12.93 &   0.03  &   2.09  &   0.14  &  5.36    &  0.05 &   2.93  &0.18 &  0.95 &  0.30    \\
    19  &    0.5476 &     8.21  &   0.05  &   3.61  &   0.10  &  11.09  &   0.07 &  7.72   & 0.40 & 9998 & 9998    \\
    20 &     0.8424 &     2.718  &   0.50 &  9998&  9998  &9999 & 9999 & 9999 & 9999 & 9999 & 9999  \\  
    21  &    0.6980 &     11.56  &   0.03  &   2.27  &   0.22  &   5.66  &   0.10  &  6.20  & 0.13 & 9998 & 9998    \\
    24  &    1.1586 &     13.61  &   0.04  &  9999&  9999  &9999 & 9999 & 9999 & 9999 & 9999 &9999   \\
    25 &     0.8094 &     6.00  &    0.04 &  9998&  9998  &9999 & 9999 & 9999 & 9999 & 9999 &9999   \\    
    26a  &   0.7027  &     3.76  &   0.01   &  1.24   &  0.30   &  2.95  &   0.05  &  22.65   &  0.01  &   5.89  &   0.05 \\
    26b  &   0.7023  &    35.75  &   0.01   &  5.32   &  0.06  &  12.85   &  0.05 &   25.34   &  0.01   &  9.03  &   0.05  \\
    28 &     0.6612  &    5.49   &   0.05  &  0.61   &   0.25   &  1.89  &   0.20 &   15.85     &  0.04 &    5.00   &  0.09    \\
    29  &    0.6619  &     2.44  &   0.09  &  0.36   &  0.40    & 1.21   &   0.20  &   1.02 & 0.26 & 9998 & 9998    \\
    31  &    0.6868  &    2.52   &   0.11  &  9999    & 9999   &  2.13    &  0.30 &   1.73  & 0.10 & 9998 & 9998    \\ 
    32   &   0.7268 &    2.25    & 0.10  & 9998 &9998  & 0.65 & 0.20 & 0.51 &0.20  & 9998 & 9998 \\ \hline
 \end{tabular} 
 \end{center}
}
\end{table*} 

\begin{table*} 
{ \begin{center}
\tiny 
 \caption{\scriptsize Measured emission line fluxes (F$_\lambda$) in unit of 10$^{-17}$ 
(ergs cm$^{-2}$ s$^{-1}$) and the errors in percent in the low-$z$ EL galaxies 
in CFRS/UDSR/UDSF fields (C/R/F) }  

\label{Fluxtablow}

\begin{tabular}{lrrrrrrrrrrrrrrrr} \hline
 Slit  &    H$\beta$ & e$_{H\beta}$\% & \ion{[O}{iii]}$_5$ & e$_{o5}$\%   & \ion{[O}{iii]}$_4$ & e$_{o4}$\% &
H$\alpha$ & e$_{H\alpha}$\%  & \ion{[N}{ii]}$_2$  &  e$_{n2}$\% & \ion{[N}{ii]}$_{1}$ & e$_{n1}$\% & \ion{[S}{ii]}$_{1}$ 
& e$_{s1}$\% &\ion{[S}{ii]}$_{2}$ & e$_{s2}$\%  \\  [1mm]
  (1) & (2) & (3) & (4) & (5) & (6) & (7) & (8) & (9) & (10) & (11) & (12) & (13) & (14) & (15)& (16)& (17) \\  \hline
  C &&&&&&&&&&&&&&&  \\
     3  &    9999      &   9999  &    1.31  &  0.19   &   9998   &    9998  &  3.45     &  0.17    &   0.54  &    0.21  & 9998    & 9998   &  9997   &   9997  & 9997  & 9997   \\ 
     9  &   2.21       &   0.11  & 0.92 & 0.18 &0.31 & 0.06 & 10.25 & 0.09 &   3.53 & 0.12 & 1.18 & 0.12 & $<$1.38 & ---  &$<$1.65 &  --- \\        
    12  &     10.09    &   0.05  &  30.90   &    0.02 &   10.97  &   0.04  & 9999   & 9999  & 9999    & 9999   & 9999    & 9999  & 9999    & 9999  & 9999    & 9999   \\
    13  &   4.49 & 0.04  &1.95 & 0.10 & 0.65 & 0.11 & 13.58 & 0.04 & 5.78 & 0.12 &1.04 & 0.12 & 3.56 & 0.10 & 1.96 & 0.19  \\
    19  &      0.77    &   0.08   &   1.44  &   0.10  &   0.43   &   0.32  & 3.41 &   0.14   &   1.37   &   0.27   & 9998    & 9998     & 9997   &    9997 &     9997 & 9997\\ 
    21  &  1.23 & 0.16 & 1.77 & 0.15 & 0.61 & 0.15  & 8.76 & 0.05 &1.80 & 0.36 & 1.06& 0.32 &   1.80 & 0.17 & 2.07 & 0.14\\
    23  &      9999     &	9999  &    9999   &   9999  &    9999  &   9999  & 25.77 &   0.01   &  37.30   &   0.01   &  11.31 & 0.03   &   5.68   &   0.08   &   5.40  &   0.12  \\
    30  &   4.76 & 0.06 & 5.43 & 0.04 &1.87 & 0.04 &17.18 & 0.04 & 3.49 & 0.12 &1.35 & 0.13 &  3.95 & 0.18 & 2.61 & 0.15 \\  \hline

 R &&&&&&&&&&&&&&&  \\
    1      &    4.83    &  0.10   &   13.68  &   0.02   &   4.60  &   0.05 &17.06 &   0.02   &   2.21   &   0.10   & 9998     & 9998    &  2.95    &  0.50   &   2.68 &0.10 \\ 
    6     &     4.62   &  0.14   &   4.31  &    0.09   &   9998   & 9998   & 19.35    &  0.03    & 12.14   &   0.08  &  3.49   &   0.10  &    6.83  &    0.15   &  4.22   &  0.40 \\ 
    11    &     4.42     & 0.05   &   3.22  &    0.08   &   0.94  &   0.06 & 13.92   &  0.03   &   8.21  &    0.03  & 9998      & 9998    &  6.16   &   0.08     & 7.19    & 0.11  \\ 
    12    &    9999      & 9999  &      9999  &   9999  &    9999    & 9999   & 11.23   &  0.03   &   6.16    &  0.07  & 9998      & 9998   &   4.42   &   0.12  &    2.73    & 0.18 \\ 
    13    &    9999      & 9999   &     9999  &   9999  &    9999    & 9999   & 4.29   &  0.08   &   2.57    &  0.10  & 9998       & 9998   &   1.49   &   0.30    &  0.70   & 0.30  \\ 
    16      &   9999     & 9999  &     9999  &    9999  &    9999    & 9999   & 24.90  &   0.01  &    5.55   &   0.04  & 9998       & 9998   &   7.62   &   0.04   &    4.81   &0.10 \\   \hline
F &&&&&&&&&&&&&&&  \\
    11   &   10.43   &    0.03   & 6.11   &   0.07   &   9998    &   9998  & 30.10  &   0.01    & 13.76    &  0.02  & 7.50       &  0.06  &  10.32    &  0.07   &   8.59  &   0.06 \\ 
    15   &    1.87   &    0.10   & 3.92   &   0.05   &   1.49   &   0.11 &   7.98 &   0.04   &   0.75  &     0.32  &  0.60    &  0.40  &    1.71   &   0.16     & 0.82 & 0.40 \\ 
    22   &    9999    &   9999    & 4.62   &   0.08    &  1.46   &   0.21  & 30.40   &  0.02    & 9.95 &    0.05   &   2.88    &  0.17   &   7.69  &   0.05     & 6.04  & 0.13  \\  \hline
 \end{tabular} 
\end{center}
}

{\tiny Notes: 
      \ion{[O}{iii]}$_5$ and o5 mean \ion{[O}{iii]} $\lambda$5007; 
      \ion{[O}{iii]}$_4$ and o4 mean \ion{[O}{iii]} $\lambda$4959; 
      \ion{[N}{ii]}$_2$ and n2 mean \ion{[N}{ii]} $\lambda$6583; 
      \ion{[N}{ii]}$_1$ and n1 mean \ion{[N}{ii]} $\lambda$6548;
      \ion{[S}{ii]}$_1$ and s1 mean \ion{[S}{ii]} $\lambda$6716;
     \ion{[S}{ii]}$_2$ and s2 mean \ion{[S}{ii]} $\lambda$6731.
The H$\gamma$ emission line fluxes of
CFRS12, 19 and UDSF15 are given in Table 4.
}
\end{table*} 

\subsection{Balmer decrement and extinction}

The extinction inside the galaxy can be derived using the decrement between 
the two Balmer lines:  H$\gamma$/H$\beta$ for our high-$z$ galaxies, and
H$\alpha$/H$\beta$ for the  low-$z$ galaxies. Case B recombination with a
density of 100\,cm$^{-3}$ and a temperature of 10\,000\,K was adopted,  the
predicted ratio is 0.466 for $I_o(H\gamma)$/$I_o(H\beta)$ and 2.87 for
$I_o(H\alpha)/I_o(H\beta)$ (Osterbrock 1989). Using the interstellar
extinction law given by Fitzpatrick (1999) with $R$=3.1 ($R=A(V)/E(B-V)$), the
extinction can be readily determined.  Using the Balmer decrement method we
find a median extinction of  $A_V$(Balmer)=1.68 (=1.82 for the $z>0.4$
sample). Extinction corrected Balmer lines (either H$\beta$ or H$\alpha$) can
be used to estimate the SFR, which could be then tested by comparison with 
SFR obtained from infrared flux.

\subsection{Comparison between SFRs deduced from mid-IR and from Balmer lines}

To compare the SFRs from IR and Balmer line, 
we adopt the calibrations
from Kennicutt (1998)  based on the
Salpeter's initial mass function (IMF) (Salpeter 1955) 
with lower and higher mass cutoffs of 0.1 and 100 $M_{\odot}$.
The median SFR$_{\rm IR}$ of our sample galaxies
is about 31 $M_{\odot}$\,yr$^{-1}$ for the $z>0.4$ galaxies (it is 
19 $M_{\odot}$\,yr$^{-1}$ when the low-$z$ galaxies are included).
The obtained median SFR$_{2.87f(H\beta)}$
is about 28 $M_{\odot}$\,yr$^{-1}$ for $z>0.4$ galaxies. For most galaxies,
SFR$_{2.87f(H\beta)}$ are consistent with SFRs estimated from infrared luminosities
(Fig.~\ref{AVSFR}(a)).

\subsection{A robust estimate of the extinction coefficient}

\begin{figure}  
 \centering
   \includegraphics[bb=76 332 238 652,width=7.4cm,clip]{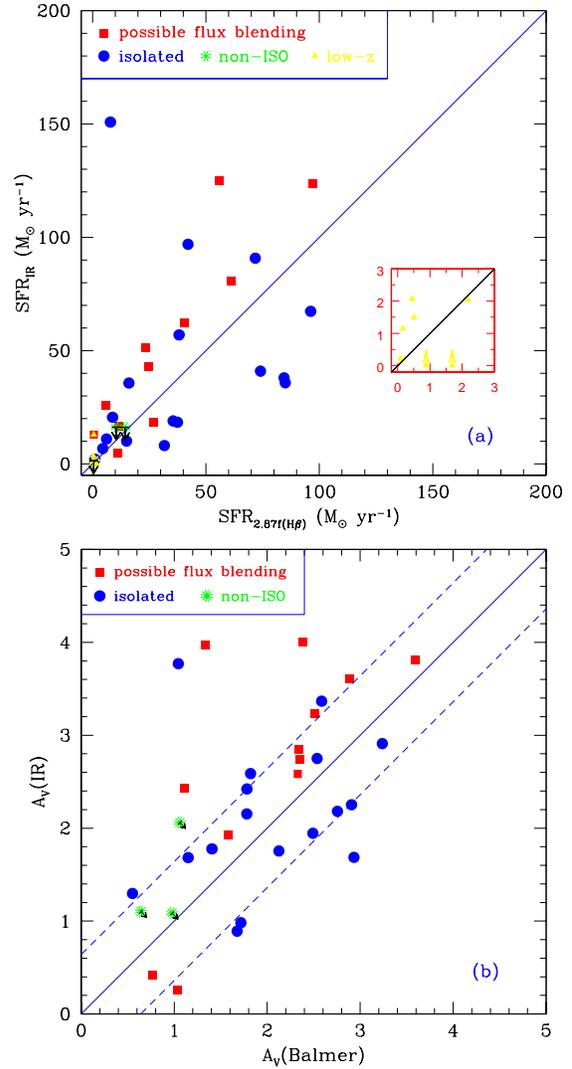}
\caption {{\bf (a)} The SFRs estimated from the extinction corrected Balmer lines 
($I_o$(H$\alpha$)$\approx$ 2.87$I_o$(H$\beta$)),
compared with the SFRs estimated from infrared fluxes. 
The small zoomed figure on the bottom right is 
for the low-$z$ sample. Relative error bars ($\Delta$SFR/SFR) are quoted in Tables 1 and 2, and average 
to 30\% (see however a discussion by Cardiel et al, 2003).
{\bf (b)} The relation between the extinction $A_V$ values derived
from Balmer decrement and by the energy balance between 
the IR radiation and the optical H$\beta$ emission line luminosities.
The two dashed lines refer to the results with $\pm$0.64 rms.
The different kinds of symbols in the two figures roughly show 
whether the infrared fluxes of the galaxies 
are affected by nearby objects according to their images from the VLT and the ISOCAM: 
 the $filled~squares$ represent the objects for which the IR fluxes are possibly 
affected by other nearby ISO objects;
the $filled~circles$ show the objects for which no contamination is expected;
the $asterisks$ mark the objects which are not detected by ISO, 
 and their $A_V$(IR) are just upper limits. 
 }
\label{AVSFR}

\end{figure}

Because of the large uncertainties related to the measurements of the
$H\gamma$ line, we need to verify the quality of our derived extinction. This
can be done assuming that the infrared data provide a robust SFR estimate
for IR-luminous galaxies  (Elbaz et al. 2002; Flores et al. 2004a).  We
estimate a new dust extinction coefficient, $A_V$(IR), by comparing the
infrared SFR with the SFR calculated from the optical H$\beta$ emission line:
the energy balance between IR and H$\beta$ luminosities.  Fig.~\ref{AVSFR}(b)
shows that  the derived $A_V$(IR) is consistent with $A_V$(Balmer) for most
galaxies, most of them falling in the $\pm$0.64 rms discrepancy. Few objects
however lie outside the $\pm$0.64 rms as shown  in Fig.~\ref{AVSFR}(b) (also
see Table 6). For three of them (filled squares, CFRS10, CFRS19 and UDSF13) 
showing $A_V$(IR) much larger than $A_{V}$(Balmer),  we believe that the
discrepancy could be related to a possible overestimate  of the IR flux due to
contamination by neighboring ISO sources  (the ``possible flux blending"
sources).

The derived median value of $A_{V}$(IR) is 2.18 ( =2.36 for $z>0.4$ galaxies)
for our sample, which is slightly larger than the extinction derived from
optical Balmer lines.  This trend might be related to the fact that  infrared
radiation includes fluxes from the optical thick \ion{H}{ii} regions, which
might be obscured to contribute to the detected optical emission lines. An
extreme example is CFRS25, CFRS03.0932, which shows $A_{V}$(Balmer)= 1.04 and
$A_{V}$(IR)= 3.77. It is an extreme edge-on disk galaxy with an inclination
of $\sim$79$^{\circ}$ (Zheng et al. 2003). Most of the optical light of  the
whole galaxy is strongly hidden by dust (screen effect), and the detected
optical Balmer lines just trace the star formation of  a few optical-thin
\ion{H}{ii} regions lying on the edge of the galaxy.

The median extinction in our sample is lower than those of the local IRAS
sample by V95 who obtained the median $E(B-V)$ (= $A_V$/3.1) values  0.99 for
the \ion{H}{ii} LIRGs, and 1.14 for the LINERs.  This could be due to the
fact that V95 only studied the  central $\sim$2\,kpc parts of the IRAS
galaxies,  which could be more affected by dust than the whole galaxy light 
as studied in distant galaxies.  The derived median extinction of our
galaxies  is comparable to that of radio-detected Sloan Digital Sky Survey 
(SDSS) galaxies
($A_{H\alpha}$=$A_{V}$/1.25=1.6,  Hopkins et al. 2003). It is much higher
than those   of the local normal star forming galaxies for which the median
$A_V$ $\approx$ 0.86  (K92 and J20). 

Comparison between the extinctions derived from IR and optical Balmer lines 
(Fig.~\ref{AVSFR}(b)) provides a significant reduction of the error bars derived from
the single H$\beta$/H$\gamma$ ratio, to $\sim$ 0.64 for $A_{V}$. This might also provide a
useful diagnostic for the dust and star formation properties in individual galaxies.
In the following, we have to adopt a reliable $A_{V}$ which describes as best as possible the global
properties of each individual galaxy. The one based on the H$\beta$/H$\gamma$ ratio shows 
a large uncertainty because it is based on the faint H$\gamma$ emission line. Diagnostic
diagrams and metal abundance determination often depend on the \ion{[O}{ii]} $\lambda$3727/H$\beta$ ratio and
then on the adopted extinction coefficient. It is uncertain whether such a ratio can be obtained as
a global parameter for a given galaxy. On the other hand, using $A_{V}$(IR) could lead to
overestimates of the extinction (and then to underestimates of the metal abundance) since it
accounts for the most obscured regions. However the good correlation found by 
Flores et al. (2004a) between SFR$_{\rm IR}$ and SFR$_{H\alpha}$ for LIRGs implies that strong obscuration are not
preponderant in the energy balance for these galaxies. Because our study of the metal abundances
is only based on starbursts and LIRGs (no ULIRGs), we adopt in the following $A_{V}$(IR) for
estimating the extinction, while paying attention to how our results would be affected if using 
$A_{V}$(Balmer).

\subsection{Continuum colors}

\begin{figure} 
   \centering
  \includegraphics[bb=70 256 238 730,width=7.5cm,clip]{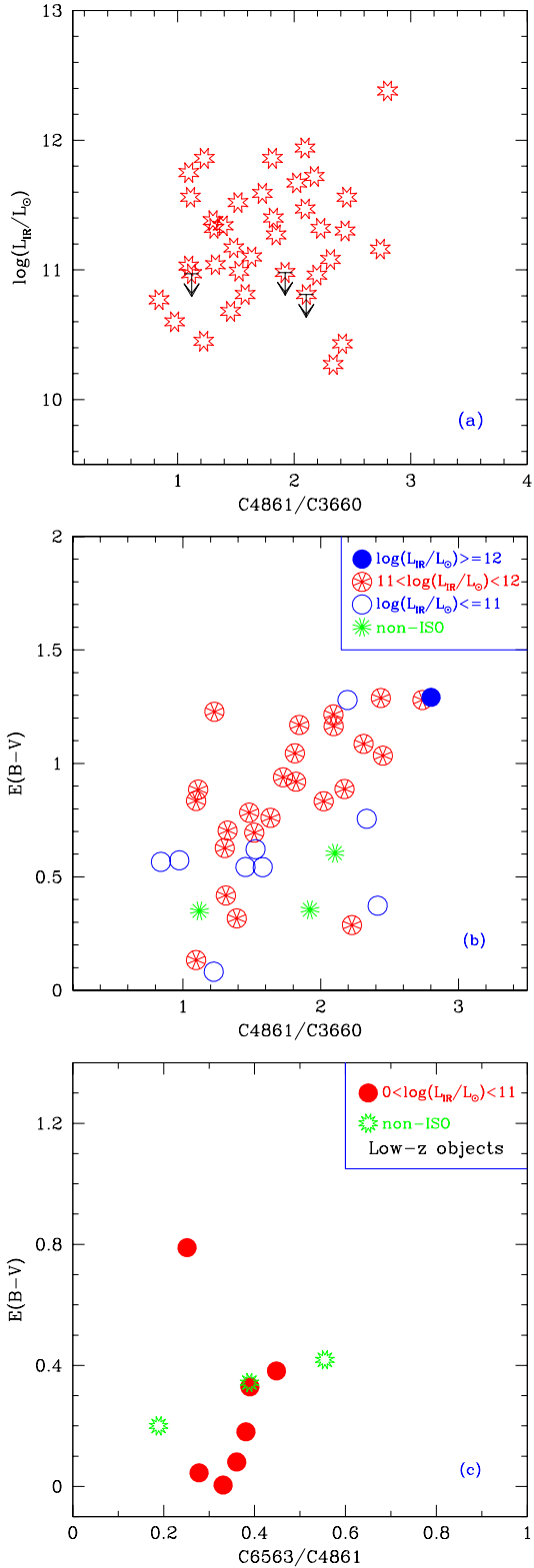}
      \caption{ The far-IR luminosities and color excesses (extinction) 
as functions of continuum colors for our sample galaxies:
{\bf (a)} IR luminosities, 
{\bf (b)} extinction for the high-$z$ sample, 
{\bf (c)} extinction for the low-$z$ sample.
The continuum colors are defined as the ratios of the continuum levels close
 to the lines (no extinction correction)  (on both sides $\sim$4\AA~around the line), 
 with a typical uncertainty 0.03.
              }
         \label{FigCou}
   \end{figure}

The continuum colors of our sample galaxies are determined by the ratios of
the continuum levels at 4861\AA$~$ and 3660\AA$~$ (C4861/C3660, for the
high-$z$ sample) and at 6563\AA$~$ and 4861\AA$~$ (C6563/C4861, for the
low-$z$ sample) (no extinction correction).  We use the same three IR
luminosity bins as V95 for log($L_{\rm IR}$/$L_{\odot}$) ($\leq11$, between
11 and 12, and $\geq$12)  in the following studies. Fig.~\ref{FigCou}(a)
shows the IR luminosity against the C4861/C3660 color for our sample
galaxies.  It seems that this plot shows that the more IR luminous galaxies with 
log($L_{\rm IR}$/$L_{\odot}$)$>11$ have redder colors than the less luminous
galaxies with  log($L_{\rm IR}$/$L_{\odot}$)$\leq 11$.  However, the
trend is very weak(less than 1$\sigma$) and similar to the result of V95
(their Fig.17, for C6563/C4861 vs. IR luminosity). Fig.~\ref{FigCou}(b) shows
the color excess (extinction)  against the C4861/C3660 color for our sample. 
It shows a weak positive correlation,  i.e., the redder color, the higher
dust extinction, which is similar to that of the local IRAS sample shown by
Fig.5   of V95 (with C6563/C4861 color). Fig.~\ref{FigCou}(c) shows the color
excess against the C6563/C4861 color for the low-$z$ sample. It seems that a
weak correlation exists as well, similar to that  of V95 (on their Fig.5).  
The median color C6563/C4861 value is about 0.4. If the dust extinction is
considered to correct the continuum color roughly, the median color  
(C6563/C4861)$_0$ (``0" means extinction correction) is about 0.35,  which is
similar to the value 0.4 obtained by V95 for their local IRAS sample  (their
Fig.\,16).

{  
\begin{table*} 
{ \begin{center} 
\scriptsize
 \caption [] { The extinction $A_V$(Balmer), $A_{V}$(IR),
the  SFR$_{H\alpha}$,``flux blending" factor ``FB", the continuum colors,
 some important emission line ratios, the oxygen abundance in ISM 
and the spectral types from the diagnostic diagrams 
(``H" for \ion{H}{ii} region galaxies, ``L" for LINERs and ``S" for Seyfert 2)
for the sample galaxies.
\ion{[O}{ii]} = \ion{[O}{ii]} $\lambda$3727, 
\ion{[O}{iii]} = \ion{[O}{iii]} $\lambda$4959 + \ion{[O}{iii]} $\lambda$5007, 
and \ion{[O}{iii]}$_5$ = \ion{[O}{iii]} $\lambda$5007. 
``FB" (Col.(4)) refers to the possible flux blending: 
``3" means ``possible flux blending",  ``1" means
``isolated", ``0" means ``non-ISO detected".  
For oxygen abundances (Col.(12)), the right up-description 
``L" means the values are determined from $R_3$ parameter. 
The uncertainties of the line ratios and metallicity are from the uncertainties of 
extinction and emission line flux measurements. 
The typical 30\% uncertainty of SFR$_{\rm IR}$ will result in a typical 0.03 dex discrepancy on the derived values.
} 

 \label{higtab}
 
\begin{tabular}{lcccccccccccl} \hline

  Slit    &  $A_V$    &SFR$_{H\alpha}$ & FB & $A_{V}$ & $\frac{C4861}{C3660}$  & log($\frac{\ion{[O}{ii]}}{\ion{[O}{iii]}_5}$) & log($\frac{\ion{[O}{ii]}}{H\beta}$) & log($\frac{\ion{[O}{iii]}}{H\beta}$) & log(R$_{23}$) & log(O$_{32}$) &
12+log($\frac{O}{H}$) & T  \\ 
        & Balmer        & $M_{\odot}$ yr$^{-1}$  &   &   IR &   &   &   &   &   &     & &   \\ 

 (1)     &$~~$(2)$~~$ & $~~$(3) & (4) & (5) & (6) & (7) & (8) & (9) & (10) & (11) & (12) &(13) \\ \hline
  C      &         &            &   &   &   &   &   &   &   &     & &   \\ 
      2  &  2.52$\pm$0.96           & 56.08$^{+58.51}_{-52.41}$      &  3    & 3.24   & 1.81      &   0.67$\pm$0.10   &  0.79$\pm$0.08  &   0.24$\pm$0.03  &  0.89$\pm$0.07 &  $-$0.54$\pm$0.10   &  8.37$\pm$0.11   &  H   \\
      6  &  2.35$\pm$0.96           & 40.35$^{+40.35}_{-42.11}$      &  3    & 2.74   & 1.11      &   0.29$\pm$0.10   &  0.47$\pm$0.09  &   0.30$\pm$0.03  &  0.69$\pm$0.05 &  $-$0.17$\pm$0.10   &  8.69$\pm$0.06   &  H \\
      8  &  2.54$\pm$1.17           & 71.75$^{+90.79}_{-67.17}$      &  1    & 2.75   & 2.17      &       ---         &   ---           &   ---            &       ---      &    ---              &      ---         &  --  \\ 
     10  &  1.11$\pm$0.93           &$~$6.05$^{+6.13}_{-4.24}$       &  3    & 2.43   & 1.48      &   1.28$\pm$0.15   &  0.56$\pm$0.07  &$-$0.60$\pm$0.13  &  0.59$\pm$0.07 &  $-$1.16$\pm$0.15   &  8.72$\pm$0.07   &  H \\
     11  &  3.59$\pm$1.40           & 97.17$^{+146.9}_{-95.20}$      &  3    & 3.81   & 1.23      &   1.20$\pm$0.14   &  0.61$\pm$0.08  &$-$0.47$\pm$0.11  &  0.64$\pm$0.07 &  $-$1.07$\pm$0.14   &  8.67$\pm$0.09   &  H\\
     12  &  1.41$\pm$0.90           &$~$4.54$^{+4.44}_{-3.55}$       &  1    & 1.78   & 0.97      &       ---         &     ---         &   0.58$\pm$0.01  &    ---         &   ---               &  8.34$\pm$0.01$^{L}$ &  H  \\
     14  &  2.75$\pm$0.90           & 35.49$^{+34.70}_{-33.71}$      &  1    & 2.18   & 1.32      &   0.73$\pm$0.07   &  0.67$\pm$0.04  &   0.06$\pm$0.04  &  0.77$\pm$0.04 &  $-$0.61$\pm$0.07   &  8.55$\pm$0.05   &  H \\
     17  &  3.24$\pm$1.48           & 96.17$^{+154.1}_{-93.30}$      &  1    & 2.91   & 1.73      &   1.52$\pm$0.20   &  0.92$\pm$0.09  &$-$0.48$\pm$0.17  &  0.93$\pm$0.09 &  $-$1.39$\pm$0.20   &  8.15$\pm$0.20   &  L \\
     19  &  1.34$\pm$1.34           &$~$0.59$^{+0.85}_{-0.45}$       &  3    & 3.97   & ---       &       ---         &     ---         &   0.32$\pm$0.05  &       ---      &      ---            &  8.57$\pm$0.05$^{L}$   &  H  \\
     25  &  1.04$^{+1.47}_{-1.04}$  &$~$7.89$^{+12.55}_{-5.35}$      &  1    & 3.77   & 2.09      &   0.50$\pm$0.15   &  0.48$\pm$0.13  &   0.10$\pm$0.05  &  0.63$\pm$0.11 &  $-$0.38$\pm$0.15   &  8.73$\pm$0.19   &  H\\
     32  &      ---                 & $~~$            ---            &  1    & 3.97   & 6.24      &   1.13$\pm$0.13   &  1.08$\pm$0.10  &   0.07$\pm$0.07  &  1.12$\pm$0.09 &  $-$1.01$\pm$0.13   &  7.78$\pm$0.25   &  L  \\
     33  &  2.59$^{+3.27}_{-2.59}$  &$~$8.83$^{+31.30}_{-8.30}$      &  1    & 3.37   & 2.31      &   1.19$\pm$0.12   &  0.89$\pm$0.09  &$-$0.18$\pm$0.07  &  0.93$\pm$0.08 &  $-$1.07$\pm$0.12   &  8.22$\pm$0.16   &  L    \\ \hline
     R   &                          &           &       &   &   &   &   &   &      &  & &   \\                                                                                                                     
      4  &       ---                & $~~$---                        &  3    & 3.97   & ---       &   0.94$\pm$0.15   &  1.32$\pm$0.13  &   0.51$\pm$0.05  &  1.38$\pm$0.11 &  $-$0.81$\pm$0.15   &  7.05$\pm$0.41   &  L  \\
      6  &  1.78$\pm$1.53           &$~$1.03$^{+1.71}_{-0.88}$       &  1    & 2.42   & ---       &        ---        &   ---           &   0.05$\pm$0.04  &        ---     &   ---               &  8.83$\pm$0.04$^{L}$   &  L?     \\
      8  &  0.55$^{+0.60}_{-0.55}$  & 16.05$^{+10.39}_{-7.23}$       &  1    & 1.30   & 1.31      &   0.52$\pm$0.06   &  0.56$\pm$0.05  &   0.16$\pm$0.02  &  0.70$\pm$0.04 &  $-$0.40$\pm$0.06   &  8.66$\pm$0.04   & H   \\
      9  &      ---                 & $~~$  ---                      &  1    & 3.99   & 2.44      &   0.39$\pm$0.10   &  1.61$\pm$0.09  &   1.35$\pm$0.03  &  1.80$\pm$0.06 &  $-$0.26$\pm$0.10   &   ---            & S  \\ 
     10  &  1.82$\pm$1.13           & 42.08$^{+51.67}_{-36.25}$      &  1    & 2.59   & 1.10      &   0.71$\pm$0.11   &  0.53$\pm$0.08  &$-$0.05$\pm$0.06  &  0.63$\pm$0.06 &  $-$0.58$\pm$0.11   &  8.71$\pm$0.07  &  H  \\
     14  &  1.72$^{+3.44}_{-1.72}$  & 84.41$^{+314.8}_{-71.30}$      &  1    & 0.98   & 1.39      &   0.45$\pm$0.48   &  0.42$\pm$0.42  &   0.09$\pm$0.07  &  0.58$\pm$0.29 &  $-$0.33$\pm$0.48   &  8.78$\pm$0.26  &  H  \\
     15  &  2.91$\pm$1.19           & 37.44$^{+48.12}_{-35.84}$      &  1    & 2.25   & 1.20      &        ---        &    ---          &   0.07$\pm$0.05  &     ---        &        ---          &  8.81$\pm$0.05  &  --       \\
     20  &  1.68$\pm$0.70           & 84.95$^{+64.50}_{-71.16}$      &  1    & 0.89   & 2.23      &   0.36$\pm$0.07   &  0.52$\pm$0.05  &   0.29$\pm$0.04  &  0.72$\pm$0.03 &  $-$0.23$\pm$0.07   &  8.65$\pm$0.04  &  H  \\
     23  &  2.49$\pm$0.84           & 74.07$^{+67.79}_{-69.11}$      &  1    & 1.94   & 1.30      &   0.75$\pm$0.08   &  0.53$\pm$0.06  &$-$0.09$\pm$0.04  &  0.62$\pm$0.05 &  $-$0.62$\pm$0.08   &  8.72$\pm$0.05  &  H  \\
     26  &       ---                &$~$  ---                        &  0    & 2.34   & 2.33      &   0.89$\pm$0.21   &  0.76$\pm$0.17  &   0.00$\pm$0.07  &  0.83$\pm$0.15 &  $-$0.76$\pm$0.21   &  8.44$\pm$0.24  &  H   \\
     32  &  2.13$\pm$1.72           & 14.98$^{+27.94}_{-13.49}$      &  1    & 1.76   & 0.84      &   0.54$\pm$0.10   &  0.73$\pm$0.08  &   0.32$\pm$0.04  &  0.87$\pm$0.06 &  $-$0.41$\pm$0.10   &  8.42$\pm$0.09  &  H  \\ \hline
     F   &                                                           &                &           &                   &                 &                  &                &                   &                   &   &   &   \\ 
      1  &  1.04$\pm$0.58           & 11.26$^{+7.07}_{-7.61}$        &  3    & 0.25   & 1.22      &   0.38$\pm$0.06   &  0.55$\pm$0.03  &   0.29$\pm$0.05  &  0.74$\pm$0.03 &  $-$0.26$\pm$0.06   &  8.63$\pm$0.03  &  H  \\
      2  &       ---                &$~$  ---                        &  1    & 1.16   & 2.71      &   0.88$\pm$0.53   &  0.68$\pm$0.42  &$-$0.08$\pm$0.23  &  0.75$\pm$0.36 &  $-$0.75$\pm$0.53   &  8.56$\pm$0.49  &  H   \\
      3  &      ---                 & $~$ ---                        &  1    & 1.87   & 2.10      &   0.54$\pm$0.19   &  0.39$\pm$0.12  &$-$0.03$\pm$0.14  &  0.53$\pm$0.09 &  $-$0.42$\pm$0.19   &  8.82$\pm$0.07  &  H   \\
      7  &  1.58$^{+2.15}_{-1.58}$  & 11.43$^{+26.58}_{-9.38}$       &  3    & 1.93   & 1.53      &   0.79$\pm$0.20   &  0.44$\pm$0.14  &$-$0.22$\pm$0.13  &  0.53$\pm$0.11 &  $-$0.66$\pm$0.20   &  8.81$\pm$0.10  &  H  \\ 
     12  &      ---                 &$~$  ---                        &  1    & 3.21   & 3.45      &       ---         &   ---           &      ---         &    ---         &    ---        &  ---                      &  --       \\
     13  &  2.39$^{+3.24}_{-2.39}$  & 27.43$^{+96.28}_{-25.37}$      &  3    & 2.39   & 2.80      &$-$0.23$\pm$0.39   &  0.86$\pm$0.34  &   1.21$\pm$0.05  &  1.37$\pm$0.15 &     0.12$\pm$0.39   &  ---            &  S  \\
     15  &  1.06$\pm$0.92           &$~~$0.51$^{+0.50}_{-0.35}$      &  0    & 2.06   & ---       &        ---        &    ---          &   0.41$\pm$0.03  & ---            &    ---              &  8.50$\pm$0.03$^{L}$ &  H       \\
     16  &  2.93$\pm$1.66           & 31.65$^{+56.95}_{-30.34}$      &  1    & 1.68   & 1.45      &   0.46$\pm$0.17   &  0.21$\pm$0.08  &$-$0.13$\pm$0.15  &  0.37$\pm$0.07 &  $-$0.33$\pm$0.17   &  8.93$\pm$0.04  &  H  \\
     17  &  1.78$^{+2.27}_{-1.78}$  & 38.18$^{+93.75}_{-32.63}$      &  1    & 2.15   & 1.52      &   1.15$\pm$0.19   &  0.69$\pm$0.15  &$-$0.33$\pm$0.08  &  0.73$\pm$0.14 &  $-$1.03$\pm$0.19   &  8.56$\pm$0.20  &  H  \\
     18  &  1.15$\pm$0.94           &$~$6.20$^{+6.32}_{-4.42}$       &  1    & 1.68   & 1.58      &   0.89$\pm$0.09   &  0.60$\pm$0.04  &$-$0.17$\pm$0.08  &  0.67$\pm$0.04 &  $-$0.77$\pm$0.09   &  8.66$\pm$0.04  &   H \\
     19  &  2.33$\pm$0.78           & 60.94$^{+51.25}_{-56.06}$      &  3    & 2.58   & 2.02      &   0.41$\pm$0.38   &  0.20$\pm$0.06  &$-$0.08$\pm$0.37  &  0.38$\pm$0.13 &  $-$0.28$\pm$0.38   &  8.92$\pm$0.08  &   H \\
     21  &  0.98$^{+1.49}_{-0.98}$  & 14.40$^{+23.22}_{-9.41}$       &  0    & 1.09   & 1.12      &   0.43$\pm$0.09   &  0.45$\pm$0.06  &   0.14$\pm$0.06  &  0.62$\pm$0.04 &  $-$0.31$\pm$0.09   &  8.74$\pm$0.04  &   H \\
     26a &  0.64$^{+1.97}_{-0.64}$  & 10.45$^{+22.28}_{-5.24}$       &  0    & 1.10   & 1.92      &$-$0.62$\pm$0.05   &  0.25$\pm$0.04  &   0.99$\pm$0.01  &  1.06$\pm$0.01 &     0.74$\pm$0.05   &  8.36$\pm$0.02  &   H \\
     26b &  0.77$\pm$0.50           & 27.11$^{+14.77}_{-15.29}$      &  3    & 0.41   & 1.10      &   0.21$\pm$0.05   &  0.50$\pm$0.04  &   0.41$\pm$0.01  &  0.76$\pm$0.02 &  $-$0.09$\pm$0.05   &  8.62$\pm$0.03  &   H \\
     28  &  2.34$^{+2.07}_{-2.07}$  & 24.60$^{+55.28}_{-22.66}$      &  3    & 2.85   & 1.82      &$-$0.04$\pm$0.19   &  0.83$\pm$0.17  &   0.99$\pm$0.03  &  1.22$\pm$0.07 &     0.16$\pm$0.19   &  ---            &   S \\
     29  &  2.88$^{+2.93}_{-2.88}$  & 23.34$^{+74.03}_{-22.32}$      &  3    & 3.61   & 2.09      &   0.91$\pm$0.25   &  0.77$\pm$0.18  &$-$0.02$\pm$0.15  &  0.84$\pm$0.16 &  $-$0.79$\pm$0.25   &  8.43$\pm$0.25  &   H \\ 
     31  &    ---                   & $~~$ ---                       &  3    & 2.36   & 1.64      &   0.51$\pm$0.29   &  0.38$\pm$0.26  &$-$0.01$\pm$0.06  &  0.53$\pm$0.18 &  $-$0.39$\pm$0.29   &  8.82$\pm$0.15  &   H  \\
     32  &    ---                   &$~~$ ---                        &  3    & 3.63   & 1.84      &   1.18$\pm$0.11   &  1.01$\pm$0.09  &$-$0.04$\pm$0.05  &  1.05$\pm$0.09 &  $-$1.05$\pm$0.11   &  7.96$\pm$0.21  &   L  \\ \hline

\end{tabular} 
\end{center}
}
\end{table*} 
}

{ 
\begin{table*} 
{ \begin{center} 
\scriptsize
{  \scriptsize
 \caption [] {The extinction $A_V$ from Balmer decrement, SFRs,
 important emission line ratios, oxygen abundance 
 in ISM and continuum colors 
 of the low-$z$ galaxies. ``nc" means no extinction correction for the emission line fluxes. 
  The uncertainties of the line ratios and metallicity are from the uncertainties of 
extinction and emission line flux measurements.
}  }

 \label{Lowtab}

\begin{tabular}{lcccccccc} \hline
 Slit  & $A_V$ & SFR  & log($\frac{\ion{[S}{ii]}}{H\alpha}$) & log($\frac{\ion{[N}{ii]}}{H\alpha}$)  & log($\frac{\ion{[O}{iii]}}{H\beta}$) &logR$_3$ & 12+log($\frac{O}{H}$) & $\frac{C6563}{C4861}$  \\  
     &             & $M_{\odot}$\,yr$^{-1}$    &      &  &  & & &   \\ 
 (1) &  (2) & (3) & (4) & (5) & (6) & (7) & (8) & (9)    \\ \hline

CFRS &             &      &      &  &  & &  &  \\
03$^{nc}$     &    ---       &     ---                  & $-$0.26$\pm$0.20     &  $-$0.80$\pm$0.12  &   ---     &  ---         & ---   &--- \\
09         &  1.30$\pm$0.38  &  0.88$_{-0.53}^{+0.24}$  & $-$0.54$\pm$0.04     &  $-$0.47$\pm$0.06  & $-$0.40$\pm$0.08   &  $-$0.55$\pm$0.11  &  9.14$\pm$0.08  & 0.5546   \\
13         &  0.14$\pm$0.15  &  0.39$_{-0.04}^{+0.04}$  & $-$0.39$\pm$0.09     &  $-$0.37$\pm$0.05  & $-$0.36$\pm$0.04   &  $-$0.49$\pm$0.06  &  9.10$\pm$0.04  & 0.2778  \\
19         &  1.18$\pm$0.52  &  0.50$_{-0.28}^{+0.18}$  &       ---          &  $-$0.40$\pm$0.13  & $~~$0.25$\pm$0.06   &   $~~$0.34$\pm$0.08  &  8.53$\pm$0.06 & 0.4482   \\ 
21         &  2.45$\pm$0.45  &  2.18$_{-1.80}^{+0.70}$  & $-$0.37$\pm$0.10     &  $-$0.69$\pm$0.15  & $~~$0.11$\pm$0.09   & $~~$0.15$\pm$0.12  &  8.66$\pm$0.08 & 0.2514  \\
23$^{nc}$  &     ---         &     ---                  & $-$0.37$\pm$0.06     & $~~$0.16$\pm$0.01  &      ---         &       ---        &       ---         &  ---  \\
30         &  0.62$\pm$0.19  &  1.69$_{-0.60}^{+0.23}$  & $-$0.42$\pm$0.10     &  $-$0.69$\pm$0.05  & $~$0.05$\pm$0.03   &  $~~$0.06$\pm$0.04  &  8.72$\pm$0.03 & 0.1889  \\ \hline
UDSR &             &         &                          &    &  & &  &  \\
01         &  0.56$\pm$0.17  &  0.11$_{-0.04}^{+0.01}$  & $-$0.52$\pm$0.07     &  $-$0.89$\pm$0.04  & $~~$0.44$\pm$0.02   & $~~$0.60$\pm$0.03  &  8.35$\pm$0.02 & 0.3810 \\
    06     &  1.02$\pm$0.38  &  0.45$_{-0.23}^{+0.12}$  & $-$0.25$\pm$0.18     &  $-$0.21$\pm$0.04  & $-$0.05$\pm$0.07   &  $-$0.07$\pm$0.09  &  8.81$\pm$0.06& 0.3896 \\
    11     &  0.25$\pm$0.16  &  0.16$_{-0.03}^{+0.02}$  & $-$0.02$\pm$0.06     &  $-$0.23$\pm$0.02  & $-$0.14$\pm$0.04   &  $-$0.19$\pm$0.05  &  8.89$\pm$0.04& 0.3608 \\
12$^{nc}$  &     ---         &       ---                & $-$0.20$\pm$0.11     &  $-$0.26$\pm$0.04      & --- &--- &---  & --- \\
13$^{nc}$  &     ---         &       ---                & $-$0.29$\pm$0.19     &  $-$0.22$\pm$0.07  & --- &--- &--- & ---  \\
16$^{nc}$  &     ---         &       ---                & $-$0.30$\pm$0.05     &  $-$0.65$\pm$0.04  & --- &--- &--- & --- \\  \hline
UDSF &             &         &                          &    &  & &  &  \\
11 &  0.01$_{-0.01}^{+0.09}$ &  0.09$_{-0.00}^{+0.01}$  & $-$0.20$\pm$0.04     &  $-$0.34$\pm$0.01  & $-$0.23$\pm$0.03   & $-$0.31$\pm$0.04   &  8.98$\pm$0.03  &  0.3308 \\
   15      &  1.06$\pm$0.29  &  0.51$_{-0.27}^{+0.10}$  & $-$0.50$\pm$0.19     &  $-$1.03$\pm$0.14  &  $~~$0.30$\pm$0.04  & $~~$0.41$\pm$0.06   &  8.48$\pm$0.04  &  0.3887  \\ 
22$^{nc}$  &     ---         &---                       & $-$0.35$\pm$0.07     &  $-$0.49$\pm$0.05      & --- &--- &---  &---   \\        \hline
\end{tabular} 
\end{center}
}
\end{table*} 
}

\section{Diagnostic Diagrams}

%
\subsection{High redshift galaxies}

The diagram of log(\ion{[O}{ii]} $\lambda$3727/H$\beta$) vs.
log(\ion{[O}{iii]} $\lambda\lambda$4959,5007/H$\beta$)  can be used to
distinguish the \ion{H}{ii} region-like objects from the LINERs and 
Seyferts. The \ion{H}{ii} region-like objects can be \ion{H}{ii} region in external
galaxies, starbursts, or \ion{H}{ii} region galaxies, objects known to be
photoionized by OB stars.

\begin{figure}
  \centering
   \includegraphics[bb=70 256 238 730,width=7.5cm,clip]{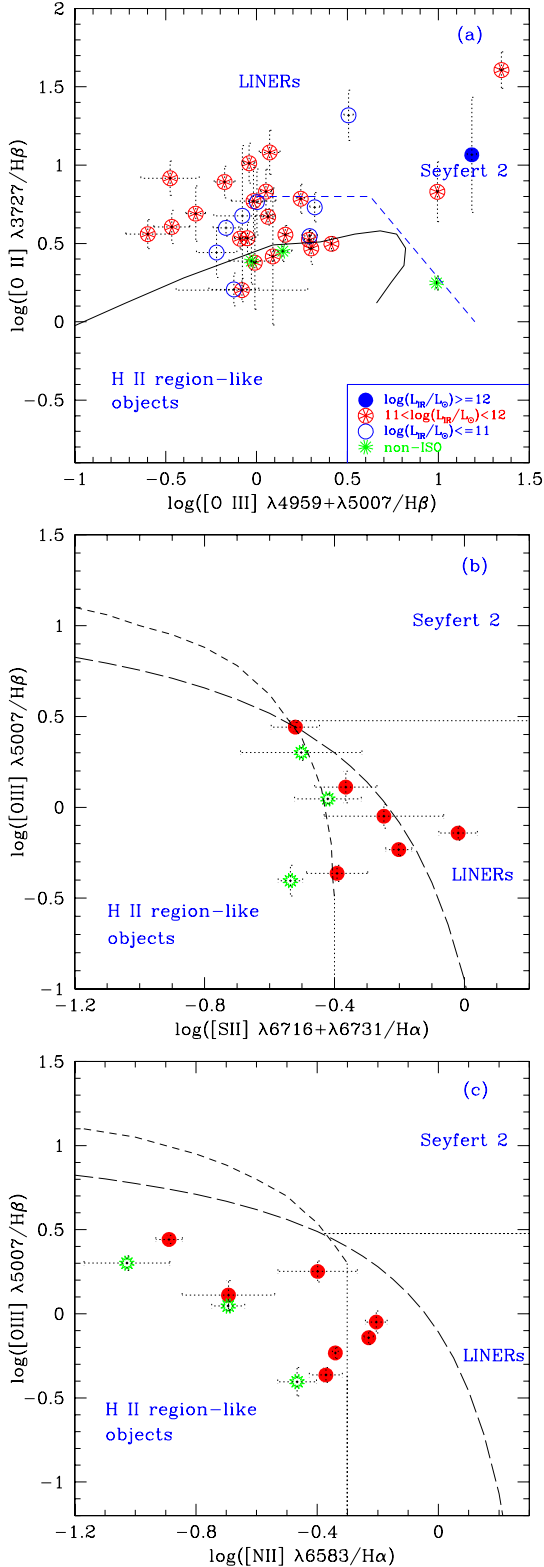}
      \caption{{\bf (a)} Diagnostic diagram for our high-$z$ sample.
The solid line shows the theoretical sequence from McCall et al.(1985),
which fits the local extragalactic
\ion{H}{ii} regions well with metallicity decreasing 
 from left to right.
{\bf (b), (c)} Diagnostic diagrams for the 
low-$z$ sample, with symbols as in Fig.~\ref{FigCou},
and the long-dashed lines
are taken from Kewley et al. (2001), others are from Osterbrock (1989).
              }
         \label{FigDig}
   \end{figure}

Figure~\ref{FigDig}(a) gives the diagnostic diagram for our $z$$>$0.4
galaxies, and shows that most of the objects are \ion{H}{ii} region galaxies,
and are consistent with the theoretical fitting of the local extragalactic 
\ion{H}{ii} regions  (the solid line, from McCall et al. 1985).  The dashed
line shows the photoionization limit for  a stellar temperature of 60,000K
and empirically  delimits the Seyfert 2 area from the \ion{H}{ii} region 
area (also see Hammer et al. 1997). From this plot, eight objects are
identified to be AGNs, including five LINERs  (CFRS17, 32, 33, UDSR04 and
UDSF32) and three Seyfert 2 galaxies (UDSR09, UDSF13, 28). An AGN fraction of
$\sim$23\% is identified from the diagnostic diagram.  However, this ratio
can be decreased to 11\% when $A_V$(Balmer) instead of $A_V$(IR) is used to
correct the emission line fluxes. One reason for this is the higher
extinction of $A_V$(IR) results in higher \ion{[O}{ii]} emission line flux
corrected by extinction,  hence, a higher LINER fraction.  The AGN fraction
is consistent with previously  reported results (17\%, Fadda et al. 2002).

\subsection{Low redshift galaxies}

For the low-$z$ galaxies, the diagnostic diagrams of 
\ion{[O}{iii]} $\lambda$5007/H$\alpha$ vs. 
\ion{[S}{ii]} $\lambda$$\lambda$6716,6731/H$\alpha$ and 
\ion{[O}{iii]} $\lambda$5007/H$\alpha$ vs. 
\ion{[N}{ii]} $\lambda$6583/H$\alpha$ 
are available to diagnose their
source of ionization 
(Veilleux \& Osterbrock 1987; Osterbrock 1989).
Fig.~\ref{FigDig}(b) and \ref{FigDig}(c) 
show these properties of the sample galaxies.

From the \ion{[S}{ii]}/H$\alpha$ vs. \ion{[O}{iii]}/H$\beta$ relations,
most of the galaxies are LINERs with low ionization levels.
However, from the \ion{[N}{ii]}/H$\alpha$ vs. \ion{[O}{iii]}/H$\beta$ relations, 
most of the galaxies are \ion{H}{ii} region galaxies since
only two objects (20\%) show the LINER character (UDSR06, 11).
Also, most of them will be classified to be "Star Forming" galaxies 
by using the corresponding diagnostic for the SDSS sample (Kauffmann et al. 2003; Brinchmann et al. 2003).
Thus, to study the diagnostic diagrams of such emission line galaxies, 
these two diagrams are needed simultaneously (also see Liang et al. 2004).
 Considering the limits given by Kewley et al. (2001) 
(the long-dashed lines on Fig.~\ref{FigDig}(b) and \ref{FigDig}(c)),
most of the low-$z$ galaxies would be classified as \ion{H}{ii} regions.
It may infer that most of them occupy a region intermediate between
 LINERs and \ion{H}{ii} regions. 

\subsection {\ion{[O}{iii]}/H$\beta$ versus \ion{[O}{ii]}/\ion{[O}{iii]} diagram for high-z
galaxies }

\begin{figure}
\centering
\input epsf
\epsfverbosetrue
\epsfxsize 7.8cm
\epsfbox{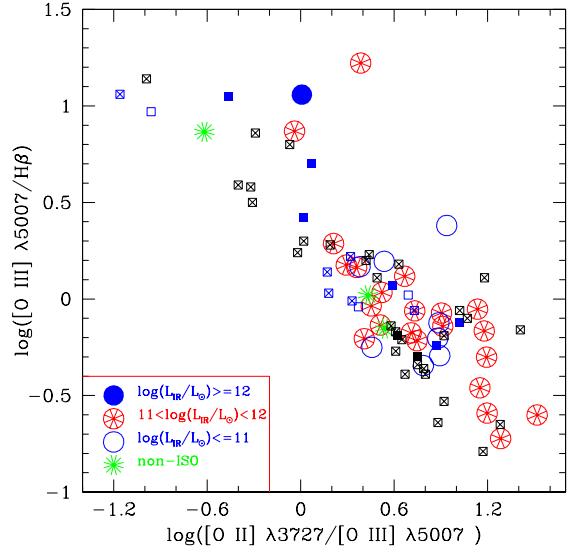}
\caption{ \ion{[O}{iii]}$\lambda$5007/H$\beta$ versus 
 \ion{[O}{ii]}$\lambda$3727/\ion{[O}{iii]}$\lambda$5007 
 emission line flux ratios of our sample galaxies, 
 compared with the local IRAS sample from V95. 
 The typical (median) uncertainties of the two parameters in logarithm are 0.08 and 0.12 dex, 
respectively. The $squares$ are the IRAS data: the $open~squares$
represent the galaxies with log($L_{\rm IR}/L_{\odot}$)$<$11, 
the $squares$ $with$ $a$ $cross$ $in$ $the$ $middle$ 
are the galaxies with 11$\leq$log($L_{\rm IR}/L_{\odot}$)$\leq$12, 
and the $filled~squares$ are the galaxies with 
log($L_{\rm IR}/L_{\odot}$)$>$12.  Most of our sample galaxies 
have relatively low ionization levels (\ion{[O}{iii]}/H$\beta$$<$3).
}
\label{V95_O23}
\end{figure}

The (\ion{[O}{ii]} $\lambda$3727/\ion{[O}{iii]} $\lambda$5007) 
emission line ratio 
follows a sequence from 
low-excitation \ion{H}{ii} regions to high-excitation \ion{H}{ii} regions
(Baldwin et al. 1981).
Fig.~\ref{V95_O23} shows the 
log(\ion{[O}{iii]} $\lambda$5007/H$\beta$) 
vs. log(\ion{[O}{ii]} $\lambda$3727/\ion{[O}{iii] $\lambda$5007}) 
diagnostic relation for our sample,
compared with that for local IRAS LIRGs of V95.
 Most of our galaxies lie in the bottom right region
indicating low ionization levels (\ion{[O}{iii]}/H$\beta$$<$3).
The apparent excess of high ionization objects in V95 could be attributed to the fact 
that they only studied the central $\sim$2kpc of the IRAS galaxies, and that they could be
more sensitive to a possible high ionization central AGN. Another possible selection bias in V95 might also
contribute to this excess, since they have been able to detect \ion{[O}{ii]} $\lambda$3727
line only in their brightest and more
distant sources, which likely include more objects with high ionization levels.

\section{Abundances in interstellar medium and luminosity-metallicity relation }

Chemical properties of gas and stars within
a galaxy are like a fossil record chronicling its history of star formation and its present 
evolutionary status. 
The high quality optical spectra from VLT/FORS2 make it possible, for the first time,
to obtain the chemical abundances in ISM for such a large sample of high-$z$ 
LIRGs.

\subsection{Metallicities estimated from diagnostic diagram}

\begin{figure}
   \centering
   \includegraphics[bb=76 332 238 652,width=7.5cm,clip]{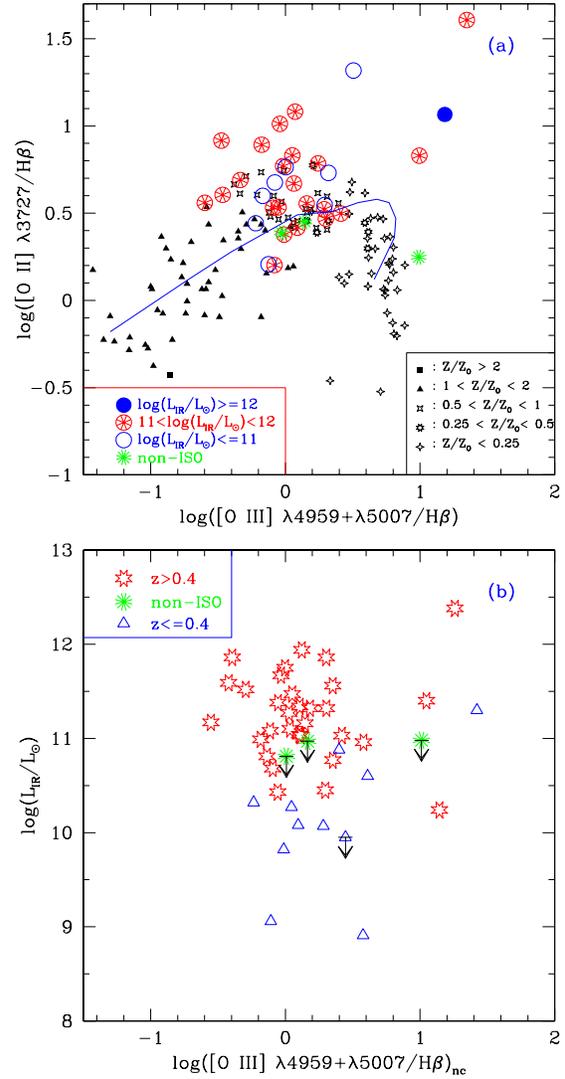}
 
  \caption {{\bf (a)} The relation between log(\ion{[O}{ii]} $\lambda{3727}$/H$\beta$) and
log(\ion{[O}{iii]} $\lambda\lambda{4959,5007}$/H$\beta$) for z$>$ 0.4 galaxies,
compared with a sample of local \ion{H}{ii} regions and \ion{H}{ii} galaxies 
with different metallicities (see Fig.12 of Hammer et al. 1997 for references of
\ion{H}{ii} samples).  The solid line shows the theoretical
sequence from McCall et al. (1985), same as in Fig.~\ref{FigDig}(a).
{\bf (b)} The relation between the infrared luminosities and 
the log(\ion{[O}{iii]} $\lambda\lambda{4959,5007}$/H$\beta$) 
(not corrected by extinction), which 
includes more data points than figure (a) panel since some galaxies can be added 
according to their \ion{[O}{iii]} to H$\beta$ ratios 
which are almost independent of extinction. 
}
         \label{OHfig1}
   \end{figure}

 On the diagnostic diagram of log(\ion{[O}{ii]} $\lambda{3727}$/H$\beta$)
vs. log(\ion{[O}{iii]} $\lambda\lambda{4959,5007}$/H$\beta$), the local
\ion{H}{ii} region samples with different metallicities lie in different
areas.  Moreover, they follow the empirical sequence from McCall et al.
(1985), which fits the local \ion{H}{ii} galaxies well with metallicity
decreasing from the left to the right (see Fig.12 of Hammer et al. 1997).
The corresponding relations for our galaxies are given in 
Fig.~\ref{OHfig1}(a) (the larger points),  together with the local
\ion{H}{ii} regions with different metallicities   (the smaller points, the
representative metallicities of the different symbols are shown in the box
on  the bottom right, $Z_0$ is the solar metallicity). The solid line shows
the theoretical sequence from McCall et al. (1985).  This diagram shows that
our high-$z$ \ion{H}{ii} region galaxies  fall in the local sample well, and 
have metallicities of 0.5$Z_{\odot}$$<$$Z$$<$$Z_{\odot}$. One non-ISO galaxy,
UDSF26a, perhaps has low metallicity with $Z$$<$0.25$Z_{\odot}$. It seems
that there is no obvious difference in metallicities between the more
luminous infrared \ion{H}{ii} galaxies (log($L_{\rm IR}/L_{\odot}$)$>$11) and
other less luminous infrared samples (log($L_{\rm IR}/L_{\odot}$)$\leq$11).

This plot shows that the horizontal-axis parameter,  log(\ion{[O}{iii]}
$\lambda\lambda{4959,5007}$/H$\beta$), can trace the metallicities of the
\ion{H}{ii} galaxies roughly, following a trend that increasing values (up to
$\sim$1.0) corresponds to lower metal abundances. Also, this ratio is almost
independent of extinction. Therefore, we further obtain the log($L_{\rm
IR}/L_{\odot}$)  vs. log(\ion{[O}{iii]} $\lambda\lambda{4959,5007}$/H$\beta$)
(no extinction correction) relation shown by Fig.~\ref{OHfig1}(b), in which
more data points are included though their H$\gamma$ emission  lines (for
extinction) and/or \ion{[O}{ii]} $\lambda$3727  shift out of the rest-frame
spectra.  Fig.~\ref{OHfig1}(b) indicates that there is almost no obvious
correlation between the two parameters, if one exists, a very weak
correlation may show the decreasing \ion{[O}{iii]}/H$\beta$ ratio following
the increasing  IR luminosity for the high-$z$ galaxies when the three
Seyfert 2  galaxies (with log(\ion{[O}{iii]}
$\lambda\lambda{4959,5007}$/H$\beta$) $>$1.0) are excluded.

\subsection{Oxygen abundances estimated from strong emission lines}

The ``direct" method to determine chemical compositions requires the electron
 temperature and the density of the emitting gas (Osterbrock 1989).
In a best-case scenario, the electron temperature of the ionized medium 
can be derived from the ratio of a higher excitation auroral line, such as 
\ion{[O}{iii]} $\lambda{4363}$ to \ion{[O}{iii]} $\lambda\lambda{4959,5007}$. 
However, \ion{[O}{iii]} $\lambda{4363}$ is too weak to be measured except in extreme
metal-poor galaxies, and becomes extremely weak in more metal-rich environments 
due to abundant heavy elements reducing collision excitation of the upper levels. 
In this case, the
oxygen abundances may be determined from the ratio of 
\ion{[O}{ii]}+\ion{[O}{iii]} to H$\beta$ lines (``strong line" method). 
The general parameter is $R_{23}$:
$
R_{23}=({\ion {[O}{ii]}\lambda{3727}+\ion {[O}{iii]}\lambda{4959}+\ion 
{[O}{iii]}\lambda{5007}})/{H\beta}.
$

Many researchers have developed formula for converting $R_{23}$
into oxygen abundance. The different calibrations have been checked
by Kobulnicky \& Zaritsky (1999)  and Kewley \& Dopita (2002). 
Recently, using the calibration given by McGaugh (1991),
Kobulnicky et al. (1999) presented two analytic formulae to
convert $R_{23}$ into 12+log(O/H) for the metal-rich and metal-poor branches,
respectively. Here we only adopt the calibration for the
metal-rich branch since the related sample includes only luminous galaxies ($M_B$$<-$18)
(Kobulnicky et al. 1999):

\begin{eqnarray}
   & & 12+log(O/H)= 12-2.939-0.2x-0.237x^2\nonumber\\
  & -& 0.305x^3-0.0283x^4- y(0.0047-0.0221x  \nonumber \\
  & -& 0.102x^2-0.0817x^3 -0.00717x^4),
\end{eqnarray}
where $x$ refers to log$R_{23}$, and $y$ refers to log$O_{32}$:
$
{\rm log}(O_{32})={\rm log}((\ion {[O}{iii]}\lambda{4959}+\ion {[O}{iii]}\lambda{5007})/\ion {[O}{ii]}\lambda{3727}).
$
To translate $R_{23}$ to be 12+log(O/H), the different calibrations  will
cause slightly different oxygen abundances. In fact, it is not important
which calibration is used here since  we are interested in the relative
abundances between local and distant galaxy samples  analyzed in the same
manner.  For the comparison samples in Fig.~\ref{MBOH1} (K92; J20;
Kobulnicky et al. 2003; Lilly et al. 2003), the same calibration (Kobulnicky
et al. 1999) has been adopted. 

Due to the limits of wavelength ranges, \ion{[O}{ii]}$\lambda{3727}$ emission lines
shift out of the visible wavelength
in the low-$z$ galaxies. Oxygen abundances in ISM of 
these galaxies can be estimated by 
$R_3$ parameter for this case (Edmunds \& Pagel 1984):
$
R_3=1.35\times (\ion {[O}{iii]}\lambda{5007}/H\beta).
$
Then, their 12+log(O/H) values can be obtained
using the empirical relation proposed by Vacca \& Conti (1992)
(also see Coziol et al. 2001):
$$
log(O/H)=-0.69\times log R_3 - 3.24, ~~~~~(-0.6\leq log R_3 \leq 1.0).
$$

The derived 12+log(O/H) values of the galaxies are given in Col.(12) of Table
6 (for high-$z$ sample) and Col.(8) of Table 7 (for low-$z$ sample), which
show that our high-$z$ \ion{H}{ii} region galaxies have 12+log(O/H) from 
8.36 to 8.93, with a median value of 8.67. Two low-$z$ galaxies have higher
metallicities than 9.0.  We should notice that here and hereafter the 
mentioned ``oxygen abundance" or ``metallicity" refers to ``gas-phase" value in the ISM.

For the solar metallicity,  Anders \& Grevesse (1989) obtained a value of
12+log(O/H)=8.90, Grevesse \& Sauval (1998) got 8.83,  whereas Allende Prieto
et al. (2001) gave a preferred solar value of 8.68. Therefore,  in the
following, our discussion is based wherever possible on 12+log(O/H) values
rather than  metallicities relative to solar, in order to avoid confusion,

\subsection{Luminosity-metallicity relation}

In the local Universe, metallicity is well correlated with the absolute
luminosity of galaxies  (Zaritsky et al. 1994; Contini et al. 2002; Melbourne
\& Salzer 2002; Lamareille et al. 2004).  Based on the current understanding
of cosmic evolution, the volume-averaged star formation rate was higher in
the past  (Madau et al. 1996; Lilly et al. 1996; Flores et al. 1999) and the
overall metallicity in the Universe  at earlier times was correspondingly
lower.  We might expect galaxies to be considerably brighter at a given
metallicity (i.e. luminosity evolution)  if they are forming stars at
higher rates. A high or intermediate redshift galaxy sample ought to be
systematically displaced from the local sample in the  L$-$Z plane if
individual galaxies reflect these cosmic evolution processes. However, if
local effects such as the gravitational potential and ``feedback" from
stellar winds and supernova regulate the star formation and chemical
enrichment process, the L$-$Z relation might be less dependent on the cosmic
epoch.  In fact, feedback could confuse the use of metallicity as a
simple metric (Garnett 2002).

\begin{figure*} 
 \centering
  \includegraphics[bb=71 360 410 646,width=16cm,clip]{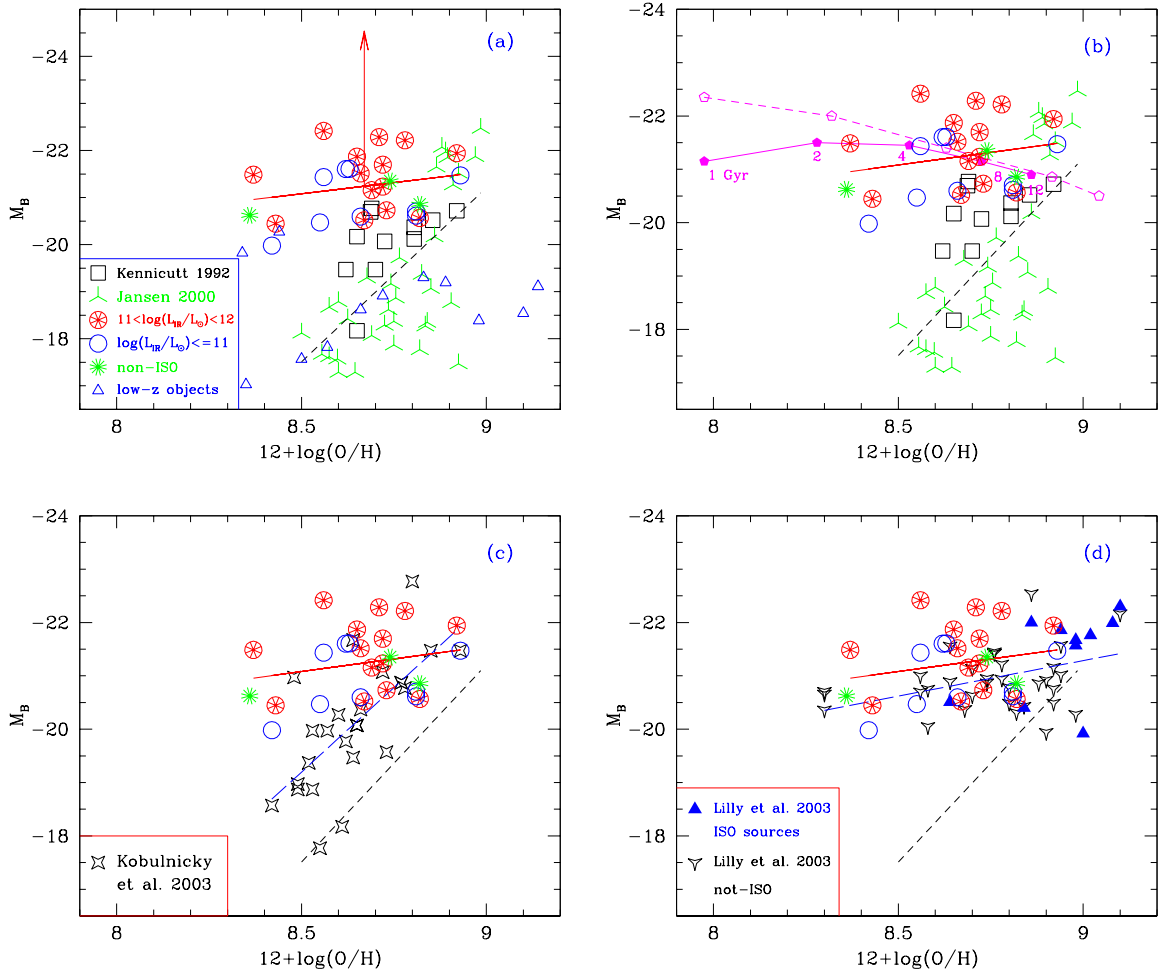}
 
\caption { 
 The $M_B$$-$metallicity relation of our distant LIRG samples 
 (with the typical uncertainty of 0.08 dex on metallicity), 
compared with other samples and Pegase2 models:
{\bf (a)} with the local galaxies from K92 and J20; the vertical arrow connecting with 
the solid fit line shows the maximal extinction effect on $M_B$, 
 assuming an average extinction
correction of $A_V$=2.36. 
{\bf (b)}  Pegase2 infall models are 
superimposed assuming a total mass of $10^{11} M_{\odot}$ and infall times 
of 5 Gyr and 1 Gyr (the solid and dashed lines with pentagons, respectively); 
the different timescales are indicated below pentagons. 
{\bf (c)} with the galaxies with 0.4$<$$z$$<$0.82 from Kobulnicky et al. (2003).
{\bf (d)} with the galaxies with 0.47$<$$z$$<$0.92 from Lilly et al. (2003); solid triangles
represent the LIRGs in the Lilly et al. sample for which we suspect that their location should
move to lower metal abundance values (see text), reconciling their results to ours and those of Kobulnicky
et al. 
The linear least-squares fits of the samples are given:
the $solid~line$ is for our distant LIRGs, 
the $short$-$dashed~line$ for the local sample,
and the $long$-$dashed~lines$ for the other two high-$z$ samples.
}

\label{MBOH1}
\end{figure*}

Figure~\ref{MBOH1} presents the $M_B$ vs. 12+log(O/H) relations 
for our LIRG sample, compared with the local (from K92 and J20) 
and the other two high-$z$ samples
(from Kobulnicky et al. 2003 and Lilly et al. 2003). 
The $M_B$ values of the compared samples have been corrected to be the same
cosmological model as ours and in the AB system.
The linear least-squares fits for the 
corresponding galaxy samples are also given 
 by considering the metallicity as an independent variable. 

\subsubsection{Comparison with local disks}

Figure~\ref{MBOH1}(a) compares the L-Z relation for the LIRGs to that of 
local disks (K92 and J20), which are restricted to moderately star forming
galaxies (EW(H$\beta$)$<$20\AA) following  Kobulnicky et al. (2003).  For the
local disks, there is a correlation between $L_{B}$ and Z,  with some
dispersion at  low luminosity, which could be related to different star
formation histories  at different  epochs. For the brightest/more abundant
disks, this might be translated  into a mass-abundance relation, assuming
that their B lights are dominated by emissions from  intermediate or old
stellar  populations. Most of the low redshift sample galaxies (the
$triangles$)  lie in the disk locus, which could  be simply related to the
fact that they show moderate SFRs, and are not so different from the local
disks. Two of the low-z galaxies (CFRS09, 13)  show high metal abundances
(12+log(O/H)$>$9.0). They are the so-called ``CFRS H$\alpha$-single" galaxies
studied by Liang et al. (2004), and are over-abundant spirals. The situation
for distant LIRGs is far more complex.  At a given metal abundance,  
all of them
show much larger B luminosities than local  disks, which corresponds to
$\delta M_B$= 2.5 mag at the median 12+log(O/H)= 8.67 (with the median
$M_B$=$-$21.24). These galaxies show $\sim$0.3 dex lower metallicity
than that of the local disks 
with the similar B luminosity (e.g. the median $M_B$=$-$21.24). 
Adopting $A_V$(Balmer) instead of $A_V$(IR) would move the
median metal abundance value of our galxies by +0.03 dex.

The small $M_B$ variation with metallicity is  probably related to
selection effect because distant ISOCAM sources likely correspond to luminous
(and massive?) systems (also see Franceschini et al. 2003). As an aside, they
are also consistent with an infall model  (single-zone Pegase2  from Fioc \&
Rocca-Volmerange 1999) as displayed in Fig.\,11 of Kobulnicky et al. (2003)
for a $10^{11} M_{\odot}$ galaxy (our Fig.~\ref{MBOH1}(b)).  The model
assumes a SFR proportional to the gas mass where the galaxy is built by
exponentially decreasing infall of primordial gas with an infall  timescale
of 5 Gyr (the solid line with pentagons).  Here the nucleosynthesis yields of
stars  (from the B-series models of Woosley \& Weaver 1995) have been
arbitrarily reduced by a factor of 2 to avoid overproduction of metals at
late times (see  Kobulnicky et al. 2003). The dispersion of the points around
this relation  may be reproduced by adding singular burst of star formation
of $10^{6-7}$ $M_{\odot}$ on the model galaxy .

However, we believe that Fig.~\ref{MBOH1}(a) does not tell us all of the
story.  Indeed, distant LIRGs show SFRs extending from 30 to several hundreds
of $M_{\odot}$\,yr$^{-1}$  and high gas  extinctions. Conversely to local
quiescent disks, their B luminosities are dominated by young  stars, and as
such, are strongly affected by dust effects.  The latter cannot be accurately
estimated from their spectral energy distribution, without a careful
modelling of stellar populations, IMF and of the dust geometry.   We can
estimate the maximal B luminosity of distant LIRGs, which can be reached if
all blue stars were embedded in  the ionized gas. This ``maximal" dust
correction is represented in Fig.~\ref{MBOH1}(a) by a big vertical arrow 
connecting with the linear least-squares fit of our high-$z$ LIRGs sample, 
assuming an average  extinction correction $A_V$(IR)=2.36 (or 3 mag at
4350\AA~).  Then, at a given metal abundance, LIRGs have B luminosities by
far larger than those of the local disks,  which excess $\delta M_B$ ranging
from 2.5 mag to more than 5 mag at the median 12+log(O/H)= 8.67. Assuming an
infall gas model, distant LIRGs could be interpreted as forming very massive
systems with  total mass ranging from $10^{11} M_{\odot}$ to $\le$ $10^{12}
M_{\odot}$,  which extend from massive disks to massive ellipticals. 
       
    It is valuable to notice that, at the given magnitude (the median
    value), the metallicity of the distant LIRGs are also lower by $\sim$0.3
    dex than those of other local samples (Contini et al. 2002; Lamareille et
    al. 2004, Melbourne \& Salzer 2002). Because the above studies are based
    on UV or H$\alpha$ emission, they mostly include low luminosity (mass?)
    systems in the local Universe. In Fig.~\ref{MBOH1}(a) we have chosen to
    compare our results to those of more massive objects, i.e. the spiral
    galaxies   from K92 and J20, because this provides us a better tool to
    understand evolutionary effects.

\subsubsection{Comparison with other high-$z$ samples}

Figure~\ref{MBOH1}(c) compares the distant LIRGs with  the high-$z$ galaxies
from Kobulnicky et al. (2003) with $0.4<z<0.82$  (EW(H$\beta$)$<$20\AA).
Kobulnicky et al. estimated the O/H values using $R_{23}$ and $O_{32}$
parameters obtained from the corresponding equivalent widths of the lines, 
which are believed to be less affected by dust extinction (see Kobulnicky \&
Phillips 2003). The metallicities of their galaxies are in similar range to
ours,  but the galaxies are fainter at a given metallicity. The median
($M_B$,12+log(O/H)) of their sample is  about (8.64,$-$20.08). The difference
between the two samples decreases at increasing metallicity, from $\delta
M_B$=2.3 mag at 12+log(O/H)$\sim$ 8.42  to $\delta M_B$= 0 at
12+log(O/H)$\sim$ 8.85. The $\delta M_B$ is $\sim$1 mag at the median
abundance of 8.67 of our LIRGs. This discrepancy in $M_B$ reflects that our
sample galaxies are brighter and possibly more massive than the rest-frame
blue selected sample  of DGSS galaxies.  

Figure~\ref{MBOH1}(d) compares the distant LIRGs with the distant CFRS sample
studied by  Lilly et al. (2003).   The $M_B$ values of the two galaxy samples
are very similar,  from about $-$19.8 to $-$22.5.  The linear least-squares
fits of the two samples  (the solid line is for our sample, and the
long-dashed line is for Lilly's sample)  show a non-significant difference in
L-Z relation: $\delta M_B$$\sim0.4$ mag. For reasons of clarity, we have
restricted the sample of Lilly et al. (2003) to the  CFRS 3$^h$ and 14$^h$
fields, which have been surveyed by ISOCAM. Among this subsample of 42
galaxies, 10 have been identified to be LIRGs by ISO  (shown as $filled
~triangles$ in Fig.~\ref{MBOH1}(d)).  The derived oxygen abundances by Lilly
et al. for these 10 ISO-galaxies show a median value of 12+log(O/H)=8.98, 
which is $\sim$0.3\,dex higher than the median value of our distant LIRG
sample.  Indeed,  Lilly et al. (2003) assumed a constant extinction of
$A_V$=1 for all their galaxies, which  strongly underestimates the average
extinction for LIRGs, and then, leads to underestimated
\ion{[O}{ii]}$\lambda{3727}/H\beta$ ratios, hence   the overestimated oxygen
abundances. This effect has been checked by us in investigating the
properties of two common galaxies in the two samples.  For CFRS02
($A_V$=3.24) and CFRS06 ($A_V$=2.74), 
Lilly et al. (2003) found the 0.5 dex
and 0.3 dex larger 12+log(O/H) values than our estimates, respectively.  
If one assumes an average $A_{V}$=2.36 for LIRGs in the Lilly et al. sample,
this would move all the corresponding points  (full triangles) towards a
lower metallicity by $\sim$ 0.3 dex. Extinction effects could thus reconcile
Lilly et al.'s results with those of Kobulnicky et al. (2003).
  

%
\section{Discussion}

We have gathered a sample of 105 ISOCAM galaxies for which we present
detailed optical spectroscopic properties. The sample,  although slightly
biased towards high IR luminosity, shows several strikingly similar
properties with the local sample of IRAS galaxies studied by  V95 and Kim et
al. (1995). This includes a similar IR luminosity distribution, continuum
color, extinction and ionizing properties. We believe that our sample
provides a good representation of distant LIRG properties. We also confirm
that, for  $>$77\% of the distant LIRGs selected by ISOCAM at 15 $\mu$m, star
formation is  responsible for most of their IR emission.\newline
 
ISO distant galaxies present a L-Z diagram strikingly different from that  of
local disks. Distant LIRGs are forming stars at very large rates, and their
L-Z diagram is almost an horizontal line reaching the local disk L-Z
correlation:  they are the systems actively building up their metal content.
At the median luminosity  ($M_{B}$= $-$21.24), their median 12+log(O/H) is
about 0.3 dex smaller than the local disk value, and even less if we correct
the $M_{B}$ value for extinction.  It is unlikely that this discrepancy is
related to our  determination of metallicity, because we have adopted the
conservative assumption that all distant  LIRGs lie on the upper branch of
the $R_{23}$ - O/H diagram, i.e. with 12+log(O/H) $\ge$ 8.35. In the
following, we investigate  the relation between distant LIRGs and
present-day  disk galaxies. 

  LIRGs can reach the local disk locus by either a progressive enrichment of
their metal content or fading. Both cases are somewhat extreme.  Single-zone
infall  models with infall time from $\tau$=1 to 5 Gyr could easily reproduce
the link between distant LIRGs, the distant large disks, and the local
massive disks (see Fig.~\ref{MBOH1}b).  These models predict the total masses
of galaxies range from  $10^{11} M_{\odot}$ to $\le$$10^{12} M_{\odot}$. The
latter mass value  comes from the maximal B band luminosity (see section
6.3.1). As noticed by  Kobulnicky et al. (2003), reducing the infall time 
(to values down to few $10^{8}$ years) would lead to   overproduction of
metal at later times.  However, we believe that simple infall models  cannot
apply during the whole history of the galaxy: several factors can prevent the
star formation from being held at  very large rates, including disk self
regulation (Silk 1997), gas consumption or small timescales  related to
merging events. 

A major problem is the uncertainty about the characteristic infall time:  if
much smaller than 1\,Gyr, distant LIRGs might fade away after the burst and
be  progenitors of low mass disks. It is likely that LIRGs correspond to
specific events of strong  star formation in galaxy history: if  associated
to mergers, such events should be rather short, within few $10^{7}$ to 
$10^{8}$ years, leading to relatively small amounts of formed stellar mass
and metals.  Indeed, several consecutive bursts are  predicted by merging
simulations.  Several minor merger events may occur in a Hubble time. 
The study of the Balmer absorption lines of these LIRGs will be presented in
a forthcoming paper (Marcillac et al. 2004), in which they will
be used to quantify the mass fraction of stars born during the
starbursts as well as the duration of these bursts.

LIRG morphologies suggest that a noticeable amount of them are intimately
linked with the population of large disks.   Using HST color maps of 34
distant LIRGs drawn from the CFRS sample,  Zheng et al. (2003) showed that
36\% of the LIRGs have disk morphologies and only 17\% are major mergers of
two galactic disks, which confirms the  preliminary study in Flores et al.
(1999). Lilly et al. (1998) have gathered a small but representative sample
of large disks  ($r_{disk}\ge$ 4 $h_{50}^{-1}$) at z$\ge$ 0.5, which appears
in number density comparable to that of local large disks. Restricting this
sample to the  two CFRS fields surveyed by ISO (3$^h$ and 14$^h$), 6 (30\%)
of the 19 large disks are LIRGs ($L_{\rm IR}$ from 3 to  18 $10^{11}$
$L_{\odot}$) detected by ISO (4 of them are indeed detected by ISO but 2 of
them are in the supplementary catalog of Flores et al. 1999; also see Zheng
et al. 2003).  If we assume that the sample of Lilly et al. (1998) is a good
representation of massive disks in the 0.5$<$z$<$1 volume, and that a large
fraction of them are experiencing  strong star forming episodes (LIRGs), this
could constraint the amount of metal/stellar mass they have formed.   The
elapsed time between $z$=1 and $z$=0.5 is $\sim$2.7 Gyr, and if 30\% of
distant disks are experiencing strong star formation episodes (LIRGs), infall
time likely averages to a value close to 1 Gyr: this would suffice to produce
an amount of metal to reach the upper metal branch of local massive disks.
  
Unfortunately, at present, we only have a few K band measurements for the
sample  galaxies studied here to estimate their stellar masses.  However,
Zheng et al. (2003) has derived rest-frame K-band luminosities ranging  from
1.4\,$10^{10} L_{\odot}$ to 2.9\,$10^{11} L_{\odot}$ for 24 of their distant
LIRGs.  The characteristic time for  doubling the stellar masses of the
distant LIRGs  ranges from $10^{8}$ to $10^{9}$ years. Since in the  very
simple model (1 Gyr infall) described above, a noticeable fraction of the gas
was  not converted into stars, this supports our view that LIRGs are related
to the formation of massive systems ($\sim$$10^{11}M_{\odot}$, mostly large
disks). On average,  z$\sim$1 large disks could double their metal content
and their stellar mass to reach the massive spiral locus at z$<$0.5.  
This is consistent with Franceschini et al. (2003), who  found that the
distant IR sources in the HDF-S are hosted by massive galaxies
($M\sim$10$^{11}$$M_{\odot}$), with  an observed median star forming activity
parameter, $M/SFR$$\sim$1Gyr.

Our result is globally in agreement with that of Franceschini et al. (2003) who 
found that the host galaxies of ISO sources are massive members of groups
with typically high rates of star formation (SFR$\sim$ 10 to 300 $M_{\odot}$\,yr$^{-1}$),
and suggested that the faint ISOCAM galaxies appear to form a composite population,
including moderately active but very massive spiral-like galaxies, and very 
luminous ongoing starbursts, in a continuous sequence.

\section{Conclusion}

A large sample (105) ISO/15$\mu$m-selected sources in three ISO deep survey fields
(CFRS 3$^{h}$, UDSR and UDSF), are studied on the basis of
their high quality VLT/FORS2 spectra and the infrared data from ISO. 
Among the 92 redshift-identified objects, 
75 (64 with $z>0.4$) are classified to be EL galaxies. 
66 (55 with $z>0.4$) objects are EL galaxies out of the 
77 ISOCAM 15 $\mu$m detected sources.
This is by far the largest sample of spectra of distant ISO galaxies.  
We present here their properties derived from the emission lines.
The redshift distribution ($z_{\rm med}$=0.587) is consistent with that of previous studies,
and some galaxies belong to a $z$$\sim$0.70 cluster (in the UDFS field) 	
or to a $z$$\sim$0.166 
galaxy group (in the UDSR field). This study provides us:\newline
- an accurate calculation of the extinction coefficient from Balmer emission lines, which
 properly account for underlying Balmer absorption ($A_V$(Balmer),
  median value of 1.82 for $z$$>$0.4 sample galaxies)\newline
- an energy balance between IR and Balmer luminosities which provides us an 
  original way to calculate an IR-based extinction coefficient 
 ($A_V$(IR), median value of 2.36 for $z$$>$0.4 galaxies)\newline
- a good correlation between the above-mentioned two extinction coefficients, 
 which, for the first time, allows us to
reliably obtain the chemical properties of distant galaxies with considering 
 interstellar extinction; this diagram could also be used
as a diagnostic of their dust geometry and star formation history\newline
- the very similar general properties of $z$$>$0.4 ISO sources to those of
the local IRAS LIRGs studied by V95, including IR luminosity distribution,
continuum color,
ionization and extinction properties, which ensures that our sample is representative of LIRGs in the
distant Universe\newline
- an extinction corrected diagnostic diagram based on 
\ion{[O}{ii]}/H$\beta$ and \ion{[O}{iii]}/H$\beta$ ratio, which confirms the small fraction      
($\sim$23\%) of Seyferts and LINERs for ISOCAM 15 $\mu$m sources 
found in other studies (Flores et al. 1999; Fadda et al. 2002; 
Elbaz et al. 2002)\newline
- a determination of O/H metallicity, which ranges from 8.36 to 8.93 in 
12+log(O/H) for the $z$$>$0.4 sample; 
no obvious trend is found
between the IR luminosity and the metallicity. \newline

This study shows that distant LIRGs draw a peculiar sequence in the $M_{B}$-O/H diagram, 
which is a broad L-Z relation at bright B absolute luminosities.
When compared to local bright disks, at constant metallicity (12+log(O/H)=8.67, median value), 
LIRGs are from 2.5 to 5 magnitudes more luminous, and at constant $M_{B}$
($M_{B}$= $-$21.24, median value), they are more than two times deficient in metal. 
These properties
can be reproduced with infall models though one has to limit the infall time to avoid
overproduction of metals in late time. 
Models predict a total mass ranging from $10^{11}$$M_{\odot}$ to
$\le$$10^{12}$$M_{\odot}$, which can be twice of the stellar masses of distant LIRGs derived by 
Zheng et al. (2003) on the basis of the K band luminosities. 
In a forthcoming paper (Hammer et al. 2004) we investigate whether LIRG properties could be related
to a recent and significant star formation in massive galaxies, including spirals.

\section*{Acknowledgments} 
 We are very grateful to Dr. Jarle Brinchmann for the very detailed and valuable 
 comments which have greatly helped us in improving this paper.
  We thank Dr. Nicolas Gruel for providing us an up-dated version of his software; 
  we are grateful to Dr. Xianzhong Zheng for his comments and suggestions.  
 We thank Dr. David Koch for his help to improve the English language in the text.
 We also thank Dr. Xiang-Ping Wu for valuable discussions 
 about the cosmological parameters.
This work has been supported by grants of the french Ministry of Education and of the
K. C WONG Education Foundation and CNRS.

\end{document}